\documentclass[a4paper, 11pt]{article}
\usepackage{my_jcap}

\usepackage{subcaption}  % in the preamble

\usepackage{natbib}
\setcitestyle{aysep={}}
\usepackage{graphicx}	% Including figure files
\usepackage{amsmath}	% Advanced maths commands
\usepackage{breqn}
\usepackage{array}

\usepackage{fontawesome5}
\usepackage{color}
\usepackage{xcolor}

\allowdisplaybreaks  % allow page breaks inside align

\usepackage{mdframed}
\newmdenv[
    linecolor=gray,            % Border color
    backgroundcolor=gray!10,    % Background color (20% gray)
    linewidth=0pt,            % Border width
    roundcorner=0pt,            % Rounded corners
    skipabove=\topsep,          % Space above the box
    skipbelow=\topsep,          % Space below the box
    innertopmargin=-5pt,         % Padding inside the box
    innerbottommargin=10pt,
    innerleftmargin=5pt,
    innerrightmargin=5pt,
    splitbottomskip=0pt,        % Space between split parts
    splittopskip=30pt,           % Space between split parts
    nobreak=false,               % Allow breaking across pages
]{mybox}

\usepackage{mathtools}

\DeclarePairedDelimiter\floor{\lfloor}{\rfloor}

\newcommand{\kms}{{\rm \, km\hskip 1pt s}\ensuremath{^{-1}}}

\newcommand{\msun}{\ensuremath{\, {\rm M}_\odot}} 
\newcommand{\zsun}{\ensuremath{\, {\rm Z}_\odot}} 
         
\newcommand{\mpc}{\ensuremath{\, {\rm Mpc}}}         
\newcommand{\gpc}{\ensuremath{\, {\rm Gpc}}}

\newcommand{\eg}{{\sl e.g.},\hskip 1pt}

\newcommand{\nhat}{\hat n}

\newcommand*{\vcenteredhbox}[1]{\begingroup
\setbox0=\hbox{#1}\parbox{\wd0}{\box0}\endgroup}

\newcommand{\OrcidID}[1]{ \href[urlcolor = red]{https://orcid.org/#1}{\textcolor{lightgray}{\faOrcid}}}
\newcommand{\OrcidIDName}[2]{\href{https://orcid.org/#1}{#2}}

\newcommand{\Rtwohc}{R_{\rm 200c}}
\newcommand{\Mtwohc}{M_{\rm 200c}}

\newcommand{\deltabcm}{\delta_{\rm bcm}}

\newcommand{\fNLLoc}{f_{\rm NL}^{\rm \,loc}}

\newcommand{\fNL}{f_{\rm NL}}

\newcommand{\kOne}{k_1}
\newcommand{\kTwo}{k_2}
\newcommand{\kThree}{k_3}

\newcommand{\Agora}{\textsc{Agora}\,}
\newcommand{\Websky}{\textsc{WebSky}\,}
\newcommand{\CR}{{\text{CR}}}

\defcitealias{Scoccimarro2012PNGs}{S12}
\defcitealias{Sohn:2023:CMBbest}{S23}
\defcitealias{Sohn:2024:Colliders}{S24}
\defcitealias{paper1}{\textsc{Paper I}}
\defcitealias{paper2}{\textsc{Paper II}}
\defcitealias{Li:2022:WebskyRadio}{Li22}
\defcitealias{Omori2022Agora}{O22}
\defcitealias{LaPosta:2025:XraytSZWL}{La25}

\defcitealias{Anbajagane:2024:MapBaryonification}{A24}
\defcitealias{Anbajagane2023Inflation}{A23}

\usepackage[T1]{fontenc}   % modern font encoding
\usepackage{lmodern}       % Latin-Modern: vectorised version of CM
\usepackage{anyfontsize}   % let NFSS scale to *any* size on demand
\title{\fontsize{19pt}{24pt}\selectfont Primordial Physics in the Nonlinear Universe: \\ 
Towards particle constraints using the Weak lensing, Thermal SZ, and X-ray fields}

\author[1, 2]{\OrcidIDName{0000-0003-3312-909X}{Dhayaa Anbajagane}
(\vcenteredhbox{\includegraphics[height=1.2\fontcharht\font`\B]{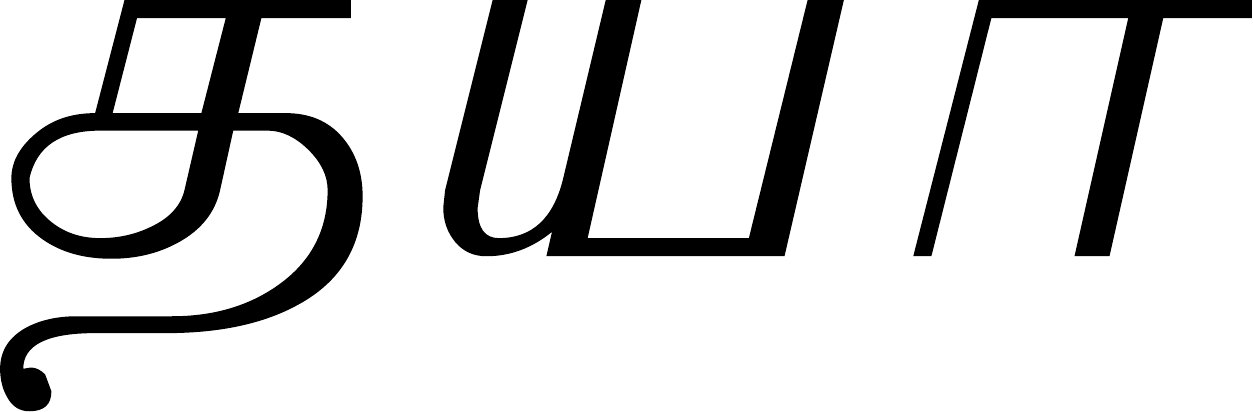}})}

\affiliation[1]{Department of Astronomy and Astrophysics, University of Chicago, Chicago, IL 60637, USA}
\affiliation[2]{Kavli Institute for Cosmological Physics, University of Chicago, Chicago, IL 60637, USA}
\emailAdd{dhayaa@uchicago.edu}

\abstract{Primordial non-Gaussianities (PNGs) are a broad class of features in the initial density field that are connected to the particle physics of the early Universe. Measuring the amplitude of these features directly constrains fundamental physics from these earliest epochs and lends insight into energy scales that cannot be probed with terrestrial experiments. Using a new class of simulation methods, we propagate these signatures to their impact on the formation of non-linear structure and quantify the constraining power in non-Gaussian summary statistics of weak lensing, thermal Sunyaev Zeldovich (tSZ), and X-ray surveys. We use semi-analytic baryon models that consistently include astrophysical effects across all these observables, and use foreground modeling approaches that explicitly fold in correlations between the various components. We find that the tSZ and X-ray fields have significant information about PNGs, and additionally can help self-calibrate a broad set of nuisance parameters/models by breaking parameter degeneracies. Using the second and third moments of the lensing, tSZ, and X-ray fields, we find a factor of 2 improvement in PNG constraints relative to using lensing alone. Larger improvements are expected when including more scales and other complementary summary statistics. Our multi-wavelength map maker can be found at \href{https://github.com/DhayaaAnbajagane/Vaanam}{github.com/DhayaaAnbajagane/Vaanam}. The simulations and software pipelines used in this analysis are publicly available.
}

\makeatletter
\def\@fpheader{\ }
\makeatother

\begin{document}

\maketitle
\flushbottom

%%%%%%%%%%%%%%%%%%%%%%%%%%%%%%%%%%%%%%%%%%%%%%%%%%

%%%%%%%%%%%%%%%%% BODY OF PAPER %%%%%%%%%%%%%%%%%%

\section{Introduction}
The physics of the earliest epochs of our Universe is rich in its phenomenology and is foundational to our understanding of the Universe's origin and evolution~\citep{Chen2010PNGReview, Achucarro2022InflationReview}. However, many aspects of this epoch remain unanswered. Observations of the Cosmic Microwave Background (CMB) provide a precise snapshot of the early Universe~\citep{Planck:2014:PNGs, Planck:2016:PNGs, Planck2020PNGs, BICEP:2021}, enabled by multiple high-accuracy experiments~\citep{Carlstrom2011, ACT:2016, Planck:2020:LegacyOverview}. Analyses of these data establish inflation --- a period where the cosmos undergoes a brief period of rapid, accelerated expansion~\citep{Guth:1981:Inflation, Linde:1982:Inflation, Guth2004Inflation} --- as the leading paradigm for the physics of the early Universe.\footnote{While our work focuses on particle physics models written down under an inflationary paradigm, the general framework we introduce is immediately applicable to any theory of the early Universe that makes concrete predictions for the statistics of the initial density field.}

In the simplest models of inflation, a single scalar field---the inflaton---is slowly rolling down a nearly flat potential while exhibiting weak self-interactions. The resulting primordial density field has fluctuations whose statistics follow that of a Gaussian random field. If we consider additional quantum fields or nontrivial inflaton interactions, the primordial density field can exhibit departures from this Gaussianity, known as primordial non-Gaussianity (PNG). The leading signatures of PNG are represented by the bispectrum $B(\kOne, \kTwo, \kThree)$, i.e., correlations among three Fourier modes of the density field, with an amplitude conventionally parameterized by $\fNL$ \citep[see][for a review]{Chen2010PNGReview}.\footnote{Some works have also considered the next-order contribution, the trispectrum \citep[\eg][]{Planck2020PNGs, Philcox:2025:PaperIII}. Simulation-based modeling of trispectrum signatures are significantly less sophisticated than those of bispectrum signatures. As a result, we focus only on the bispectrum, which already has a wide and rich phenomenology.}

PNGs can be characterized according to the structure of their three-point correlations, which in turn are sourced by the particle physics underlying a given model. Some models' amplitudes peak in the limit when two distances in the three-point correlation are far smaller than the third (the ``squeezed limit''), while others peak when the three distances are equivalent to each other (``equilateral limit''). However, a much broader landscape of possible correlations arise from considering additional interactions of the inflaton field with other fields present during this epoch~\citep{Chen:2010:QSF,Baumann:2011nk,Assassi:2012zq,Nima:2015:Colliders}. Since inflation may have occurred at energy scales as high as $10^{15}\,{\rm GeV}$, such interactions could involve particles, or couplings, far too energetic to be probed in terrestrial experiments. This makes the inflationary epoch a unique laboratory for probing fundamental physics at otherwise inaccessible energies.

Over the last two decades, our best approach for constraining PNGs has been the Cosmic Microwave Background (CMB). Constraints from the WMAP \citep{Hinshaw2013WMAP9} and \textit{Planck} \citep{Planck:2020:LegacyOverview} space missions have significantly furthered our understanding about the physics of the earliest epochs. Ongoing efforts have elevated the spatial correlations of the galaxy positions to a quickly growing, competitive probe of such PNG physics \citep{Cabass2022SingleFieldBOSS, Cabass2022MultifieldBOSS, Philcox2022BossPNG, Damico2022BossPNG}, with the potential to significantly supersede the CMB measurements in the near future \citep{Achucarro2022InflationReview}. This two-probe landscape of inflation is in stark contrast to analyses of our fiducial Lambda Cold Dark Matter (LCDM) cosmology, where $\mathcal{O}(10)$ probes inform our constraints \citep[\eg][]{DiValentino:2025:Tensions}. 

The key reason for this difference in the abundance of probes is the requirements for accurately modeling structure formation --- in LCDM, a large set of probes (such as weak gravitational lensing, cluster correlations, dwarf galaxy abundances, strong lensing, etc.) are viable through the availability of simulations for testing, modeling, and calibrating their analyses \citep[\eg][]{Nadler:2024:dwarf, To:2025:Y6modelling}. Such simulations accurately resolve the gravitational (non-linear) dynamics of structure formation and constitute precise predictions for the statistics of structure in our Universe \citep{Angulo2022SimReview}. Simpler analytic models are not sufficient for such modeling. Given PNG analyses rely almost exclusively on analytic models, they are not able to use ``nonlinear'' probes.\footnote{There is an active area of ``hybrid'' techniques that utilize simulations but while still using analytic, theoretically motivated formalisms \citep{Modi2020HybridEFT, Kokron2022HybridEFT}. Such techniques also explicitly require the availability of simulations.}

Most simulations of PNGs pertain to one specific PNG template (referred to as ``Local''-type PNG), with another four or so templates also being simulated by various groups \citep[\eg][]{Scoccimarro2012PNGs, Fondi:2025:Gengars}. To simulate PNGs, one must construct an initial state of the density field that exhibits the correct three-point (or higher-point) correlations. This task requires solving a high-dimensional convolutional integral \citep{Scoccimarro2012PNGs, paper1, paper2}, but this can be dramatically simplified if the PNG bispectrum can be represented in a factorizable manner, i.e. $B(\kOne, \kTwo, \kThree) = f(\kOne)g(\kTwo)h(\kThree)$. Most templates are not manifestly factorizable, and so the vast majority of the over 200 PNG signatures constrained by the \textit{Planck} team \citep{Planck:2014:PNGs, Planck:2016:PNGs, Planck2020PNGs} have not been simulated. However, in \citet[][henceforth, \citetalias{paper1}]{paper1} and \citet[][henceforth, \citetalias{paper2}]{paper2} we introduced new algorithms that efficiently handle this factorizability limitation by approximating any bispectrum with a finite set of factorizable terms. Using this method, we explored one sparsely considered observational probe of PNGs --- weak gravitational lensing (WL), which is the distortion of light from galaxies due to intervening matter distributions \citep{Bartelmann2001WLReview} --- and showed it has significant potential in contributing to the current landscape of PNG constraints. A key insight of these results, following from previous forays into the topic \citep{Marian2011fNL, Shirasaki2012fNL}, is the sensitivity of WL to peaks of the density field (which are tightly connected to cluster-scale halos) and the connection PNGs have to those super-structures \citep{Dalal2008ScaleDependentBias}.

In this work, we extend on our previous efforts by characterizing the PNG information in WL, in addition to two other fields whose signals are also known to be highly dominated by massive halos: the thermal Sunyaev-Zeldovich (SZ) field \citep{Carlstrom2002SZReview}, and the X-ray photon count-rate field \citep{Brandt:2022:CXRB}. Together, the three fields span a fairly wide range in sensitivity across halo mass, redshift, and length scale. Their joint analysis has proven useful in calibrating the effective impact of baryon evolution \citep{Ferreiera:2024:Xray, Pandey:2025:ACTDES, LaPosta:2025:XraytSZWL, Siegel:2025:Baryons} --- which has a significant impact on WL statistics \citep[\eg][]{Chisari2018BaryonsPk, Amon2022S8Baryons, Secco2022Shear, DECADE5} --- and so our approach here enables the opportunity to simultaneously constrain PNGs while marginalizing over baryons. We also ensure the signal contaminants/foregrounds for all three fields --- such as intrinsic alignments, cosmic infrared background, radio emission, active galactic nuclei (AGN), etc. --- are consistently correlated with one another. While prior work has independently used either WL data or SZ data to forecast PNG constraining power \citep{Shirasaki2012fNL, Hilbert2012fNL, Marian2011fNL, Hill:2013:tSZPNG}, this work provides a multi-wavelength, joint characterization of inflationary signals across WL, tSZ, and X-ray measurements, and serves as a guidepost towards maximizing PNG information extraction with existing or near-term datasets.

This paper is structured as follows: in Section \ref{sec:data}, we detail the simulation suite used in this work, including the various PNG models we focus on and the summary statistics we employ. Section \ref{sec:mockmaps} describes our semi-analytic approach to modeling observables, the survey configuration we model for, and a brief overview of the choices made in our forward modeling pipeline. A detailed description of the latter is provided in Appendix \ref{appx:model}. Section \ref{sec:results} then systematically details the constraining power in WL, in SZ, and in Xray. We conclude in Section \ref{sec:conclusions}. Appendices \ref{appx:LCDM_wCDM}, \ref{appx:numconvergence}, and \ref{appx:poisson} detail our LCDM and wCDM constraints, the numerical convergence of our results, and our method for correlated Poisson sampling, respectively.

\section{Datasets}\label{sec:data}

We first detail the raw data products used for this work. Section \ref{sec:data:sims} presents the N-body simulations that serve as the starting point of our modeling, and Section \ref{sec:data:models} details the different inflation models we consider (and thereby simulate). The choice of summary statistics, to compress the simulated maps into a lower-dimension space, is discussed in Section \ref{sec:data:stats}.

\subsection{Simulations}\label{sec:data:sims}

Our modeling pipeline is heavily reliant on N-body simulations, which numerically solve for the full, non-linear evolution of the density field under gravity.\footnote{All simulations assume General Relativity and do not consider any modifications to gravity.} This simulation-based approach presents one key advantage over perturbation theory-based analyses of cosmic structure \citep[\eg][]{Philcox2022BossPNG, Cabass2022MultifieldBOSS, Cabass2022SingleFieldBOSS, Damico2022BossPNG}, where the uncertainty in (analytically) modeling gravitational non-linearities is the key limitation for the latter. Through simulations, we precisely model these non-linearities. Our simulations come from the \textsc{Ulagam} suite \citep{Anbajagane2023Inflation, paper1, paper2}, which contain full-sky lightcones built using the \textsc{PkdGrav3} $N$-body code \citep{Potter2017Pkdgrav3}. \citet{Schneider2016Convergence} show that different N-body solvers cause shifts of $\leq 1\%$ for the scales we consider in this work ($k \leq 1 \mpc/h$).

The \textsc{Ulagam} suite is designed for simulation-based fisher forecasts of widefield surveys, as it has over 2000 full-sky maps at a fiducial cosmology, and between 100 to 200 full-sky maps where a single model parameter is shifted by $\pm\Delta x$. The former is critical for estimating robust covariance matrices of our chosen data vectors, while the latter is similarly critical for precise numerical derivative estimates of said data vectors (see Appendix \ref{appx:numconvergence} for our numerical convergence results). All simulations are run with $512^3$ particles, within volumes of $V = L^3 = (1 \gpc/h)^3$, where $h = H_0/(100 \kms/\mpc)$ is the dimensionless Hubble constant. The initial conditions (IC) are generated using 2nd-order Lagrangian Perturbation Theory \citep[2LPT,][]{Crocce2006LPT}. For simulations with primordial signatures, the ICs are generated using the \textsc{Aarambam}\footnote{\href{https://github.com/DhayaaAnbajagane/Aarambam}{github.com/DhayaaAnbajagane/Aarambam}} codebase presented in \citetalias{paper1}. Detailed validations of our IC methodology are presented in that work as well.

Each simulation contains 90 snapshots between $z = 10$ and $z = 0$, and \textsc{PkdGrav3} also internally produces lightcone density shells at the redshift of each snapshot, duplicating the 3D volumes as so that the observable volume extends to the one encompassed within a given redshift. Such \textsc{PkdGrav3}-generated lightcones have been used extensively in simulation-based analyses of weak-lensing \citep[\eg][]{Fluri2019DeepLearningKIDS, Fluri2022wCDMKIDS, Gatti:2024:WPH, Jeffrey:2025:Likelihood, Prat:2025:Homology}. Halos are identified in each snapshot using both a position-based Friends-of-Friends algorithm internal to \textsc{PkdGrav3} as well as by explicitly running \textsc{Rockstar} \citep{Behroozi2013Rockstar} on the snapshots. Throughout this work, we use the halo mass/radius definition $\Mtwohc = 200 \rho_c \times 4\pi \Rtwohc^3$, where $\rho_c$ is the critical density at a given epoch.

Appendix A of \citet[][henceforth, \citetalias{Anbajagane2023Inflation}]{Anbajagane2023Inflation} provides a validation of the simulation pipeline. In particular, they show the power spectrum of the matter density field agrees with those from higher-resolution simulations, such as those predicted by the Euclid Emulator \citep{Euclid2019Emu}, within $5\%$ up to $k \sim 1 h/\mpc$ at $z = 0$. The poorer resolution of our simulations results in a loss of power (i.e., the coarse resolution effectively smooths out small-scale power; see the right panel of Figure 4 in \citet{Schneider2016Convergence}) resulting in a suppression of signal. Therefore, increasing the resolution would only amplify our signal further rather than degrading it. So we do not expect any artificial boosts to our Fisher information estimates due to the lower resolution of our simulations. \citetalias{Anbajagane2023Inflation} (see their Figure 12 and Figure 13) also show that the cluster-scale halos produced in these simulations agree well with halo abundance and halo bias measurements from a wide variety of previous works.

Given the resolution of these simulations, the particle mass is $M_p = 7 \times 10^{11} \msun/h$, and halos of $M > 10^{14} \msun/h$ are resolved with at least 150 particles. Notably, we have no access to halos of group and Milky way-mass scales. In this work, the halo catalog is primarily used for modeling astrophysical effects, as detailed in Section \ref{sec:mockmaps}. The lack of lower-mass halos means we will underestimate astrophysical signals, such as the tSZ, X-ray, etc. This is a minor consideration for the tSZ model, where the emission (for the scales accessible at our map resolution) is dominated by halos above our lower-mass limit \citep[][see their Figure 7]{Battaglia:2012:Gamma}. It is more relevant for the X-ray field, which is sensitive to halos down to $\Mtwohc > 10^{13.5}\msun/h$ \citep[][see their Figure 3]{McDonald:2026:XrayFlamingo}. The astrophysics-induced corrections to the matter distribution will also be underestimated, though we expect the contribution of objects below the cluster mass-scale to be minor \citep[][see the top-right panel of their Figure 4]{To:2024:Baryons}.\footnote{Our work uses the lensing convergence, which has the same $k$-dependence as the $\xi_+$ summary statistic of the spin-2 shear field, and not of the $\xi_-$ statistic. The latter probes even smaller scales, and therefore has an increased dependence on baryons, and other small-scale effects \citep{Bartelmann2001WLReview}.} Thus, the impact of resolution on our halo catalog is generally a $\mathcal{O}(10\%)$ effect or less in all three fields. While cross-correlations can shift the mass sensitivity of the modeling, the difference is generally insufficient to alter our discussion above \citepalias[][see their Figure 9]{Anbajagane:2024:MapBaryonification}.

\subsection{Inflationary Models} \label{sec:data:models}

Our primary goal is to characterize the PNG constraining power provided by various survey datasets. We consider six different inflationary models in this work, distinguished by the particle physics interactions they arise from. As mentioned before, these models can be characterized by their bispectra, $B(\kOne, \kTwo, \kThree)$. It is also common to present the models through their shape function, $S = (\kOne\kTwo\kThree)^2 B(\kOne, \kTwo, \kThree)$.

We start with two popular templates from the literature, the Local \citep{Komatsu:2001:fnlLoc} and Equilateral \citep{Senatore2010WMAP5pngs}, which represent multi-field inflation and cubic interactions from single-field inflation, respectively. Almost all inflationary analyses in late-time structure have focused on the Local type as it has a unique scale-dependent signature in the halo bias \citep{Dalal2008ScaleDependentBias}, and as initial conditions with Local-type bispectra can be generated entirely in real-space without any convolutional integrals. The two templates are given as,
\begin{mybox}
\begin{align}
     S^{\rm local} & = \frac{\kThree^2}{\kOne\kTwo}+ {\rm 2~ perm.}\,,\\
     S^{\rm equil} & = \bigg(\frac{\kThree^2}{\kOne\kTwo}+ {\rm 2~ perms.}\bigg) - 2 + \bigg(\frac{\kOne}{\kThree} + {\rm 5~ perms.}\bigg)
\end{align}
\end{mybox}
where the permutations are over the exchange $k_i \leftrightarrow k_j$.

\paragraph{Quasi-Single Field (QSF) Inflation.} This model captures the impact of massive (but still relatively light) particles during inflation, These particles are characterized by the inequality $m/H_\phi \leq 3/2$, where $m$ is the particle mass and $H_\phi$ is the Hubble scale during inflation. We use $\phi$ to denote the inflaton field, and $\sigma$ to denote the additional field. The QSF shape is generated by a cubic interaction, $\sigma^3$, alongside a linear mixing, $\dot\phi\sigma$, between the inflaton and the additional field. We use the standard template defined by~\citet{Chen:2010:QSF},
\begin{mybox}    
\vskip 10pt
\begin{equation} \label{eqn:template:qsf}
    S^{\rm QSF} = 3\sqrt{3\kappa} \,\frac{Y_\nu(8\kappa)}{Y_{\nu}({8}/{27})}\,,   
\end{equation}
\end{mybox}
\noindent where $Y_\nu$ is the Bessel function of the second kind, $\kappa=k_1k_2k_3/k_T^3$, $k_T = \kOne + \kTwo + \kThree$, and $\nu = \sqrt{9/4 - (m/H_\phi)^2}$.

\paragraph{Scalar I (SI).} Another class of exchange diagrams involves a single exchange of a heavy scalar, arising from a cubic interaction of the form $(\partial\phi)^2\sigma$. Here, $\phi$ is the inflaton, while $\sigma$ is the additional scalar field. Due to the breaking of time translation symmetry during inflation, we can separately consider spatial/time-derivatives, leading to two types of interactions: $\dot{\phi}^2\sigma$ and $(\partial_i\phi)^2\sigma$. We will refer to these as Scalar I and II cases, respectively.
The bispectra generated by these interactions have been computed using cosmological bootstrap techniques~\citep{Arkani-Hamed:2018kmz,Baumann:2019oyu,Pimentel:2022fsc,Jazayeri:2022kjy}, and 
\citet{Sohn:2024:Colliders} provide the following template that approximates the exact shape:
\begin{mybox}
{\small
\begin{align} \label{eqn:template:scalar_I}
    S_{\rm col.}^{\rm SI} & = \frac{2k_1k_2k_3}{\beta k_T^2k_{12}}\bigg[ 1+ \frac{4k_3}{ (\beta+2) k_{12}} + \frac{(\beta+4)k_3^2}{ (\beta+2)^2 k_{12}^2}\bigg] \bigg(\frac{k_T}{k_{12}}\bigg)^{-\frac{\beta}{\beta+2}}\nonumber\\
    & - \frac{k_1k_2k_3}{6\cosh(\pi\mu)k_{12}^3}\bigg[{2(2\mu^4-1)}\bigg( \frac{k_3^2}{k_T^2}+ \frac{k_3(4k_T-k_3)}{k_T^2} \log \bigg(\frac{k_T}{k_{12}}\bigg)  +\log^2\bigg(\frac{k_T}{k_{12}}\bigg) \bigg) \nonumber\\
    & - \mu^2 \bigg(\frac{k_3}{k_{12}}\bigg)^{\frac{1+16\mu^2}{1+8\mu^2}} \bigg(\frac{k_3(6k_T-k_3+8\mu^2(8k_T-k_3))}{(1+8\mu^2)k_T^2} + \frac{2(3+68\mu^2+384\mu^4)}{(1+8\mu^2)^2}\log \bigg(\frac{k_T}{k_{12}}\bigg)
    \bigg)\bigg]\nonumber\\
    & + \frac{k_1k_2}{k_{12}^2} \sqrt{\frac{\pi^3\beta(\beta+2)}{\mu \sinh(2\pi \mu) }} \bigg(\frac{k_3}{k_{12}}\bigg)^{\frac{1}{2}} \cos\bigg[\mu\log\bigg(\frac{k_3}{2k_{12}}\bigg) + \delta \bigg] + 2~{\rm perm.} \,,  
\end{align}
}
\end{mybox}
where $k_{ij}= k_i+k_j$, $\beta = \mu^2 + 1/4$, and $\delta = \arg [\Gamma({\frac{5}{2}}+i\mu)\Gamma(-i\mu) (1+i\sinh\pi\mu)]$. We define $\mu \equiv \sqrt{m^2/H_\phi^2 - 9/4}$. This template consists of a non-oscillating piece and an oscillating term. The latter is of comparable amplitude to the former for low particle masses ($m/H_\phi \lesssim 3/2$) but becomes exponentially suppressed for high masses ($m/H_\phi \gg 3/2$).
The oscillatory frequency is set by the mass of the particle in Hubble units, and is larger for more massive particles. In the limit of $m/H_\phi \gg 3/2$, the above template approaches the Equilateral shape due to the exponential suppression of the oscillations.

\paragraph{Scalar II (SII).} This case parallels the Scalar I scenario above, but for the interaction $(\partial_i \phi)^2 \sigma$. 
In practice, the actual Scalar II template we use is defined as a linear combination of two terms, 
\begin{equation} \label{eqn:template:scalar_II}
    S_{\rm col.}^{\rm SII} = - S_{\rm col.}^{\rm SI} - \frac{14}{100} S_{\rm col.}^{\rm sp}\,,
\end{equation}
where $S_{\rm col.}^{\rm SI}$ is given in Eq.~\eqref{eqn:template:scalar_I} and the second term ($S_{\rm col.}^{\rm sp}$), corresponding to the actual interaction $(\partial_i \phi)^2 \sigma$, is given as in \citet{Sohn:2024:Colliders}. The expression is lengthy, and we refer interested readers to Equation D1 in \citetalias{paper1}. The linear combination decorrelates the Scalar I and Scalar II into their orthogonal set, and is analogous to a similar procedure used to define the Equilateral and Orthogonal templates from the two, correlated cubic interactions in single-field inflation \citep{Senatore2010WMAP5pngs}.

\paragraph{Non-Bunch-Davies vaccuum (NBD2).} Bispectra can also be generated if the ground state of the inflaton is excited, which is also referred to as a NBD vacuum \citep[\eg][]{Chen:2007:single-field, Holman:2008:excited-initial, Meerburg:2009:initial-state}. We consider one of the NBD models studied in \citet{Planck:2014:PNGs, Planck:2016:PNGs, Planck2020PNGs}. Following the \textit{Planck} nomenclature, we explore the ``NBD mode 2'' model,
\begin{mybox}    
\begin{align} \label{eqn:template:NBD}
    B^{\rm NBD2}(k_1,k_2,k_3) & = \frac{1}{(\kOne\kTwo\kThree)^3}\bigg[(\kTwo\kThree)^2\times \frac{1 - \cos[(\kTwo + \kThree - \kOne)/k_c]}{\kTwo + \kThree - \kOne} + 2~{\rm perm.}\bigg]\,,
\end{align}
\end{mybox}
where the excitations are generated at conformal time $\tau_c$ and set on a scale $k_c = -(\tau_c c_s)^{-1}$, with $c_s$ being the speed of sound of the fluctuations. Note that this template does not diverge in the folded limit $\kTwo + \kThree - \kOne \rightarrow 0$ due to the $1 - \cos(\cdots)$ factor in the numerator; it is exactly zero at the folded configuration, but can exhibit a large (oscillatory) amplitude in its vicinity when $k_c$ is small.

\citetalias{paper1} and \citetalias{paper2} considered a large variety of different PNG models. For simplicity, this work considers the six models above. As shown in Figure 5 in \citetalias{paper1} and Figure 3 in \citetalias{paper2}, the signatures from each model are fairly correlated\footnote{While this correlation was determined by propogating the signatures to the space of WL summary statistics, we expect the behavior to hold true even if we add the tSZ and X-ray observables as done in this work, given the latter two are highly correlated with WL.} (with some exceptions), so characterizing a subset of PNG models still provides useful guidance on the future constraints in this wider space of models.

\subsection{Statistics \& Datavector choices}\label{sec:data:stats}

In cosmology, summary statistics compress a high-dimension map/dataset into a set of low-dimension numbers that capture the relevant physics in the data. In all current analyses of PNGs, the choice of summary is limited to those that can be analytically predicted from the available models. The use of simulations gives us flexibility in choosing any statistic of interest, since we can simply forward model predictions for any given statistic and are not limited to analytic pipelines. This means, we also have flexibility in deciding how to best summarize the survey maps into a set of statistics to use in the inference process. The WL community has spent significant efforts in detailing the information content of various statistics \citep{Jain1998LensingPDF, Peel2018Moments, Cheng:2024:HSC_WST, Halder2021Integrated3ptShear, Gatti2022MomentsDESY3, Anbajagane2023CDFs, Gomes:2025:Map3, Jeffrey:2025:Likelihood, Prat:2025:Homology}, and some works have also quantified the impact of different modeling and observational systematics on these statistics \citep[\eg][]{Gatti2020Moments, Secco2022MassAp, Anbajagane2023CDFs, Gatti:2024:LFIValidation, Prat:2025:Homology, Gomes:2025:Map3, Gomes:2026:TATT}. Similar efforts have also been made in the tSZ community as well \citep[\eg][]{Wilson:2012:tSZSkew, Munshi:2013:tSZSkewSpectra, Hill:2014:PDF, Sabyr:2025:tSZHOS}.

While a variety of choices exist, and have been applied to WL data, we utilize the second and third moments of the field as our statistics \citep[\eg][]{VanWaerbeke2013CFHTLens, Petri2015MomentsMinkowski, Chang2018MassMap, Peel2018Moments, Gatti2022MomentsDESY3, Gatti:2024:LFIResults}. These moments have the advantage of being easily connected to perturbation theory models (on sufficiently linear length scales) while still probing non-Gaussian information in the field \citep{Gatti2022MomentsDESY3}. While we do not use a perturbation theory approach in this work, the connection of the moments to analytical models is a benefit in the interpretation of their information content. In the case of WL, using additional static beyond the moments has shown improvements to parameter constraints that mimic doubling the entire data volume \citep[\eg][]{Gatti:2024:LFIResults, Prat:2025:Homology}. One may reasonably expect similar behaviors for the tSZ and X-ray fields; though, even larger improvements are possible given the tSZ and X-ray field are far more dominated by massive halos and therefore, far more non-Gaussian.

Our choice of statistic, the field moments, are computed as
\begin{equation}\label{eqn:Moments}
    \langle A^{(1)}A^{(2)}\ldots A^{(N)}\rangle(\theta) = \frac{1}{N_{\rm pix} - 1}\sum_{i=1}^{N_{\rm pix}}A^{(1)}_iA^{(2)}_i\ldots A^{(N)}_i\,,
\end{equation}
for some set of fields $A^{(1)}, A^{(2)}, \ldots , A^{(N)}$, where all fields are smoothed on some scale, $\theta$. We follow previous works \citep[\eg][]{Gatti2020Moments} in performing the smoothing using a tophat-filter, which is defined in harmonic space as
\begin{equation}\label{eqn:TophatFilter}
    B(\ell) = 2 \frac{J_1(\ell\theta)}{\ell\theta}
\end{equation}
where $J_1(x)$ is a Bessel function of the first order. For $N = 2$, the moments capture the same information content as a power spectrum, and this has been checked extensively in many recent works \citep[\eg][]{Anbajagane2023CDFs, Gatti:2024:LFIValidation}. We measure the moments on 10 scales\footnote{It is also possible to define cross-moments between fields smoothed on different scales. This dramatically increases the length of the data-vector, so we have not pursued this.} spaced logarithmically between $3.2\arcmin < \theta < 200 \arcmin$, following \citet{Gatti2020Moments}.

While we could compute any number of moments with $N > 3$, \citet{Anbajagane2023CDFs} show that for lensing data from the Dark Energy Survey \citep[DES,][]{DES2005} Year 3 analysis, there is only little signal-to-noise in the fourth moment and none in the fifth moment. Recent work has since shown that the use of the third moment provides a factor of two improvement over just the second moment alone \citep{Gomes:2025:Map3, Sugiyama:2025:Map3}, and others have found that extending to the fourth moment shows little-to-no appreciable benefit to cosmology constraints from WL \citep[][see their Figure 7]{Silvestre:2025:Map4}. We treat the choice of $N \in [2, 3]$ as adequate for increasing information while minimizing numerical inaccuracy challenges caused by extending the length of our data vector. We will, however, present one analysis variant that includes a sparse subset of the fourth moments; specifically, we include moments of the form $\langle m^4_i\rangle$, $\langle m_i m_j m^2_{\rm Xray}\rangle$, $\langle m_i m_j m^2_{\rm tSZ}\rangle$, and $\langle m^\alpha_{\rm Xray} m^\beta_{\rm tSZ}\rangle$. Here $m_i$ can be any map: the tomographic lensing maps, the tSZ map, or the X-ray map. The above combination explicitly avoids considering the large set of cross-moments between just the lensing maps, while still retaining some potentially critical information from the tSZ and X-ray maps. A more thorough analysis will require far larger simulation suites than what is available in our work.

It is worth noting that for the specific case of inflationary constraints, there may be more optimal statistics beyond these aperture moments. Given that the PNG signal affects the tails of the density distribution, it is better recast into the statistics of peaks and voids \citep{Marian2011fNL, Shirasaki2012fNL, Anbajagane2023Inflation}. Thus, one might anticipate that topographical statistics --- which have already been used in WL \citep[\eg][]{Prat:2025:Homology} --- may have higher signal-to-noise for probing inflationary features in particular. We leave such studies to future work.

\section{Multi-wavelength synthetic skies}\label{sec:mockmaps}

The observables we focus on are sensitive to non-linear gravitational evolution, and therefore require simulations to accurately model. We naturally require a realistic forward model of the sky as observed in a given dataset. For the purposes of this work, we are interested in accurately capturing any component that contributes appreciably to the cosmological signal or to the noise, as these two terms determine the Fisher information (Equation \eqref{eqn:Fisher}). There have been at least one or more forward modeling efforts within various lensing, millimeter-wave (mm-wave), and X-ray surveys \citep[\eg][]{Sehgal:2010:sims, Derose:2019:Buzzard, Seppi:2022:DigitalTwin, Omori2022Agora, Ge:2025:SPTMUSE}. Of these, the \Agora simulation \citepalias{Omori2022Agora} jointly models both tSZ and lensing fields. We build on their efforts, and include the X-ray field. We now list our modeling approach, which borrows heavily from the past decade of work on this topic. Our implementation of the various operations required in this modeling are made available in the \textsc{Vaanam} codebase.\footnote{\href{https://github.com/DhayaaAnbajagane/Vaanam}{github.com/DhayaaAnbajagane/Vaanam}}

The key simulation inputs to all steps below are (i) the halo catalogs at 90 redshifts between $0 < z < 10$, and (ii) the density shells at those corresponding redshifts. The latter is estimated from the particle count maps as $\delta^i = N^{i}_{\rm p}/\langle N_{\rm p} \rangle - 1$, where $N^{i}_{\rm p}$ are the particle counts in a given (Healpix) pixel on the sky. While \textsc{PkdGrav} only provides us with 3D halo catalogs, we project them into lightcones using the same box duplication methodology used to generate the lightcones. We explicitly validated this procedure for use in \citet[][henceforth, \citetalias{Anbajagane:2024:MapBaryonification}]{Anbajagane:2024:MapBaryonification} by computing various auto/cross-correlations of the halo lightcones and the density shells. All input/output maps in this work use $\texttt{NSIDE}=1024$, corresponding to a pixel resolution of $3.2 \arcmin$. We detail the impact of this resolution choice in Section \ref{sec:results:future}.

\subsection{Semi-analytical models of astrophysics}\label{sec:mockmaps:SAMs}

First, we discuss the philosophy of our approach. Multi-wavelength analyses of the kind pursued here require a prescription for modeling astrophysical effects --- such as energetic feedback, gas cooling, etc. --- that affect all three fields we consider. In the context of a WL-only analysis, one can take three broad classes of approaches. First, we can apply scale cuts to the datavector\footnote{At the map-level, these can be trivially implemented by smoothing the maps on a sufficiently large scale \citep[\eg][]{Cheng2021WeakLensingST, Gatti:2024:LFIResults,Jeffrey:2025:Likelihood, Thomsen:2026:SBI}} and reduce sensitivity to astrophysical effects, thereby allowing us to ignore such effects in the analysis. Second, one can marginalize over simple non-physical prescriptions such as an exponential suppression of the matter power spectrum above a given wavenumber scale, or a Principle-Component Vector built from simulations \citep[\eg][]{Eifler:2015:Baryons}. Third, we can do this marginalization but using a physically motivated, halo-based model for the astrophysical corrections to the matter power spectrum \citep[\eg][]{Schneider:2015:Baryons, Arico:2021:Bacco}. For WL analyses, all three approaches can be well-motivated. However, when multiple observables have sensitivity to the gas distribution and its thermodynamic properties, only the third approach is viable as we must explicitly account for the correlated impact of astrophysics across all observables of interest. 

There are many approaches to modeling such astrophysical effects, of which the most principled are hydrodynamical simulations. These simulations solve a large set of coupled differential equations that propagate astrophysical effects into various observables \citep[see][for a review]{Vogelsberger2020Hydro}. A limitation in this approach is that one must specify the exact system of differential equations, and do so for processes we understand very little about, e.g. micro-physical processes pertaining to AGN feedback. Furthermore, there is a wide dynamical range between cosmological scales ($\sim 10^2 - 10^4 \mpc$) and astrophysical ones ($\gtrsim {\rm pc}$) so one must also provide effective, or ``subgrid,'' models of these processes, that integrate out small-scale behaviors and track only the larger scale features. There are multiple, equally justified choices of subgrid models, and these lead to a variety of predictions even for simple volume-integrated halo properties like mass fractions, velocity dispersions, etc. \citep[\eg][]{Anbajagane2020StellarProp, Lim2021GasProp, Lee2022rSZ, Cui2022GIZMO, Stiskalek2022TNGHorizon, Anbajagane2022Baryons, Anbajagane2022GalaxyVelBias, Shao2022Baryons, Gebhardt2023CamelsAGNSN}. For cosmological analyses, where we must marginalize over the space of possible astrophysical effects/signatures, it is challenging to utilize hydrodynamical simulations. Such analyses must not only marginalize over parameters in these subgrid models, but also over the possible space of subgrid models given there is significant uncertainty in the choice of model. The latter is an ill-defined question, and challenging to incorporate into practical analyses.

An alternative then is to build semi-analytic, phenomenological prescriptions that work directly in the space of observables and do not require specifying a system of equations. In the WL community, these are often referred to as ``baryonification'' prescriptions \citep{Schneider2019Baryonification, Giri2021Baryon, Arico:2021:Bacco, Anbajagane:2024:MapBaryonification}, though we will refer to them more generally as baryon semi-analytic models (SAMs). To apply such a prescription to WL, one specifies halo density profiles --- as a function of mass and redshift --- in both a dark matter-only (DMO) universe, and a dark matter-baryon (DMB) universe. This can then be used to generate a displacement function that perturbs the positions of particles in a DMO simulation to mimic astrophysical impacts on the matter distribution. The exact nature of these impacts is encoded in the choice of profiles in the DMB case. These SAM approaches are flexible and effective \citep{Schneider2019Baryonification, Giri2021Baryon}, and are particularly advantageous as they connect observable quantities like baryon fractions and halo profiles to the cosmic density distribution \citep{Giri2021Baryon, Schneider:2025:Baryonification}. There is no specification of subgrid physics in this process.

While the original displacement function formalism was defined only for 3D particle snapshots, it was then modified to work directly on projected full-sky density shells \citep{Fluri2018DeepLearning, Anbajagane:2024:MapBaryonification}, which dramatically reduced the memory usage, storage footprint, and the computation cost. Follow-up work then connected this model to predictions of the thermal gas pressure, and therefore the thermal SZ effect \citep{Anbajagane:2024:MapBaryonification, Pandey:2025:Godmax}. This allows for the self-consistent modeling of thermodynamic fields while accounting for correlated impacts of baryon evolution on the density field. In this work, we extend this existing model to include X-ray observables, as discussed further below and in Appendix \ref{appx:model:xray}. 

A detailed description of the algorithms and halo profile modeling can be found in \citetalias{Anbajagane:2024:MapBaryonification} (see their Section 2). In short, we model the DMO halo profile as a simple Navarro-Frenk-White \citep[NFW,][]{Navarro1997NFWProfile} profile with a concentration-mass relation from \citet{Diemer2019concentrations}. For the DMB case, the profile is split into a gas, star, and ``collisionless'' matter (dark matter plus satellite galaxies) component. The gas profile is a generalized NFW profile \citep{Nagai:2007:GNFW}, the star profile is a power-law with an exponential cutoff, and the collisionless matter profile is modeled as an NFW profile with an adiabatic relaxation correction that is connected to the gas and star profiles \citep{Schneider2019Baryonification}. Simple power-law or sigmoid functions characterize the trend of the gas and star mass fractions with halo mass. The default values we use are set by \citet{Giri2021Baryon} and their empirical fit to various datasets.

Finally, we note that in all discussions to follow, the field $\delta(z, \nhat)$ is estimated from N-body simulations that do not include any astrophysical modeling, and the field $\deltabcm(z, \nhat)$ is the density after including baryon corrections. All semi-analytic modeling in this work uses the \textsc{BaryonForge}\footnote{\href{https://github.com/DhayaaAnbajagane/BaryonForge}{github.com/DhayaaAnbajagane/BaryonForge}} codebase \citepalias{Anbajagane:2024:MapBaryonification}.

\subsection{Survey configurations}\label{sec:mockmaps:surveys}

Our work models a combination of widefield surveys in the optical, mm-wave, and X-ray wavelengths. We now detail the characteristics of each survey included in our forward model, as well as any comparisons to existing or upcoming datasets.

\textbf{Weak lensing:} We model the Year 10 dataset from the Vera C. Rubin Observatory \citep[henceforth, LSST,][]{Ivezic:2019:LSST}. The lensing sample is expected to cover 14,000 $\deg^2$ at a number density of $n = 30$ galaxies per sq. arcmin, with five tomographic redshifts bin that together span $0 < z < 2.5$. The $n(z)$ per bin is taken from the LSST Science Requirements Document \citep{LSST2018SRD}; see Figure 1 of \citetalias{Anbajagane2023Inflation} for a visual. While we focus on LSST Year 10, we note the potential for pathfinder analyses using the DES Year 6 \citep{Yamamoto2025} and Dark Energy Camera All Data Everywhere (DECADE) datasets \citep{DECADE1, DECADE5}. The latter has comparable number density to the Year 3 DES data \citep{Gatti2021ShearCatalog} and naturally a similar redshift distribution \citep{DECADE2}, but includes an additional 9000 $\deg^2$ of high-quality lensing data. The combination of DES and DECADE has provided some of our most precise lensing-based constraints to date \citep{DECADE4, DECADE5}. The combined data will cover the same area as LSST Year 10 but with about one-fifth the number density, so the constraints would be roughly one-third to one-fifth as constraining.

\textbf{Thermal SZ:} We focus on the Simons Observatory \citep[SO,][]{Simons:2019:Experiment} which is expected to overlap completely with area covered by the LSST cosmology sample. Note that we consider the enhanced SO configuration discussed in \citet{Abitbol:2025:SO}.\footnote{This is enabled by an early upgrade to the Large Aperture Telescope \citep{Abitbol:2025:SO} and by optimal performance of the detectors \citep{Sierra:2025:SODetector}.} The data span a frequency range of $f \in [30, 40, 95, 150, 220, 270] \, {\rm GHz}$, and we supplement it with data from the \textit{Planck} satellite mission \citep{Planck:2020:LegacyOverview}, spanning $f \in [100, 143, 217]\, {\rm GHz}$. The beam properties of the different frequencies follow Table A2 in \citetalias{Anbajagane:2024:MapBaryonification}, while the noise levels are updated to match those in \citet{Abitbol:2025:SO}, set to be achieved by 2034. We could also use mm-wave maps from the South Pole Telescope 3rd Generation \citep[SPT-3G,][]{Benson2014SPT3G}. SPT-3G have produced the lowest-noise wide-field maps of the mm-wave sky \citep{Quan:2026:SPT3G} with larger volumes of data currently being processed. While the main SPT-3G field spans 1500 $\deg^2$, the extended footprint spans 10,000 $\deg^2$ \citep{Prabhu:2024:SPT} of the sky that also overlaps with LSST. We expect our analysis here to deliver broadly similar Fisher information when swapping between the SO and SPT-3G tSZ maps, given we limit our data to $\ell_{\rm max} = 2048$, where the modes are strongly informed by \textit{Planck} data because SO and SPT-3G (as ground-based surveys) are limited by atmospheric noise on such scales. \citet{Prabhu:2024:SPT} show the constraining power of the current SPT-3G extended data will be similar to that from the first few years of SO.

\textbf{X-ray:} We focus on the eROSITA, or extended ROentgen Survey with an Imaging Telescope Array, dataset \citep[][]{Merloni:2012:eROSITA, Predehl:2020:eROSITA} and in particular the upcoming eROSITA All Sky Survey (eRASS) maps of the photon count rate in the [0.5, 2] keV energy band. While only the DR1 maps are currently available, it is expected that DR3 will contain 4 to 5 times more exposure time relative to DR1, so we utilize the DR1 maps\footnote{\url{https://erosita.mpe.mpg.de/dr1/AllSkySurveyData_dr1/HalfSkyMaps_dr1}} and scale them by a factor of 4.5. We follow \citet{Lau:2025:eRASS} and utilize the on-axis response matrix and effective area function for Telescope Module 0 (TMO).\footnote{\url{https://erosita.mpe.mpg.de/dr1/eSASS4DR1/eSASS4DR1_arfrmf}} Unlike the previous case, eRASS does not fully overlap with LSST and SO, and is instead limited to just the western ecliptic hemisphere. Using the DR1 maps, we estimate 25\% of the LSST sample will be missing X-ray data overlap. We fold in this limitation in our map-making, as is shown in Figure \ref{fig:Maps}.

\subsection{Generating mock maps} \label{sec:mockmaps:mocks}

\begin{figure}
    \centering
    \includegraphics[width=\columnwidth]{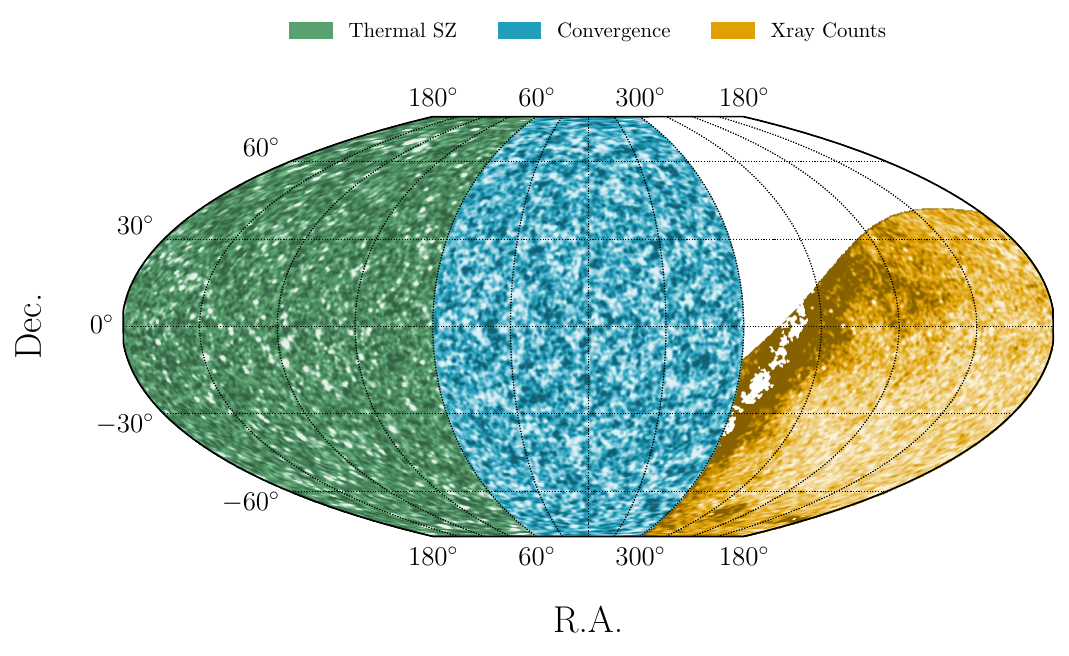}
    \caption{An example visualization of the noisy maps of the tSZ, WL, and X-ray fields, smoothed on $25\arcmin$ scales. \textit{Left:} The tSZ map is noise-dominated with clear peaks from near-by clusters (white patches), whose signal is more visually apparent due to the larger angular smoothing. \textit{Middle:} The convergence map from the third redshift bin for an LSST Y10 sample. The white (dark blue) patches denote overdensities (underdensities) in the projected matter field, and appear in roughly equal abundance. \textit{Right:} The X-ray map from eROSITA. The northern 25\% of the map is removed in our setup as a similar fraction of the LSST sample is non-overlapping with eRASS. The X-ray field has a strong feature from attenuation by the neutral hydrogen in the Milky Way, as well as in the Magellanic Clouds.}
    \label{fig:Maps}
\end{figure}

Having defined the survey configurations, we present a general overview of the modeling choices for the WL, tSZ, and Xray fields. The interested reader can find a detailed description in Appendix \ref{appx:model}, as well as in \citetalias{Anbajagane:2024:MapBaryonification}.

\textbf{Weak lensing:} The lensing convergence, $\kappa$ --- which is a line-of-sight integral of the density field --- is our main lensing observable. We prefer the scalar $\kappa$ rather than the spin-2 shear field, $\gamma^{1, 2}$, as the implementation of higher-order statistics is far simpler on a scalar field. Note that the two ($\kappa$, and $\gamma^{1, 2}$) can be linearly converted between each other; see Equation \ref{eqn:Kappa2Shear}). The convergence can directly be modeled by a weighted sum of the density shells across redshift, with the appropriate definition of these redshift-dependent weights; see Equation \eqref{eqn:convergence_definition}. 

We also include a number of other contributions in our simulation-based model of $\kappa$: (i) the intrinsic alignments of galaxy shapes/shears following the Non-linear Linear Alignment (NLA) prescription \citep[\eg][]{Troxel2015IAReview}, (ii) reduced shear, (iii) source clustering using the prescription of \citet{Gatti2023SC}, (iv) source magnification, and (v) baryon evolution using the map-level implementation of the baryon SAM \citep{Schneider2019Baryonification, Fluri2019DeepLearningKIDS, Anbajagane:2024:MapBaryonification}. In practice, we first include astrophysical effects in the density shells, and then use those modified shells to model the terms listed in (i)-(iv). Other contributions such as shape noise and masking are all included in our model, as detailed in Appendix \ref{appx:model:WL}.

\textbf{Thermal SZ:} Our key observable is the noisy Compton-y map of the sky, which is simply a line-of-sight integral of the thermal electron pressure field (and noise/foreground components). Given a baryon model, we can predict the thermal gas pressure as discussed in Appendix \ref{appx:model:tSZ}, and account for non-thermal contributions (e.g. turbulence, cosmic rays, magnetic fields, etc.) by including the appropriate model flexibility. The thermal gas pressure can be converted to electron pressure using the cosmic hydrogen and helium abundances (see Appendix \ref{appx:model:tSZ} for discussions on possible extensions of this). Once the electron pressure profiles are defined, we can paint tSZ emission around halos in our lightcone catalog and accumulate their contributions into a healpix map. In practice, we generate maps of the observed sky brightness at each frequency band, and combine them (as is done in the observational data) using a Linear Combination (LC) algorithm that maximizes the tSZ signal-to-noise in the maps \citep[\eg][]{Matt:2020:tSZACTDR4, Bleem:2022:tSZ}. Thus, our forward model will include multiple contributions to the mm-wave sky beyond just the tSZ. This includes the CMB temperature primaries, kinematic SZ effect, the Cosmic Infrared Background (CIB), radio source emissions, atmospheric noise (for ground-based survey data), and detector noise.

The emission from radio sources and the CIB dominate the small-scale power at the lower and higher frequency channels, respectively. Both sources of emission are tightly connected to galaxy formation and are therefore correlated with the cosmic density field \citep{Kashlinsky:2005:CIB}. We explicitly include this correlation by forward modeling these components from the density shells of the simulation; see Appendix \ref{appx:model:tSZ}. We broadly follow the approaches of the \Agora \citep[][henceforth \citetalias{Omori2022Agora}]{Omori2022Agora} and \Websky \citep[][henceforth \citetalias{Li:2022:WebskyRadio}]{Li:2022:WebskyRadio} simulations for the CIB and radio, respectively. The amplitude of the CIB field is also rescaled by a nuisance parameter.

\textbf{X-ray}: Our key observable is the count-rate of photons, within a given energy band, across the sky. We follow existing halo-model approaches for this observable \citep{Lau:2023:XrayPk, Ferreiera:2024:Xray, LaPosta:2025:XraytSZWL} but translate them to our simulation-based modeling approach. The X-ray signal is constructed of emission from galaxy clusters, from AGNs, and from X-ray binaries. We model the first term using the same approach as the tSZ, where we construct photon count-rate profiles for every halo in our lightcone, and generate a healpix map of their accumulated contributions. The X-ray modeling predominantly requires the same ingredients as the tSZ modeling (gas pressure or temperature, electron number density, etc.) but also requires a metallicity profile, as the photon emissivity rate depends on the chemical composition of the gas. We detail our choices for this new component in Appendix \ref{appx:model:Xray:Z}.

We then include contributions from AGNs, which could be at the 10\% level or more \citep[\eg][]{Ferreiera:2024:Xray, LaPosta:2025:XraytSZWL}, and from X-ray binaries residing in galaxies. We use data-constrained luminosity function models for AGNs and for galaxy-hosts of X-ray binaries \citep{Aird:2015:AGNXLF}, and then draw Poisson realizations of these samples that are consistent with the density shells in our simulations. Thus, all components of the X-ray sky are explicitly tied to the density field. The amplitude of the AGN component is also rescaled by a free nuisance parameter to capture uncertainties in the overall normalization of the model. For all three components, we account for relevant observational effects such as the effective area of the eROSITA telescope modules, the energy response matrix of the detectors, and the exposure time maps of the survey. See Equation \eqref{eqn:Xray:CRcluster} and Appendix \ref{appx:model:xray} for more details.

\textbf{In summary,} we follow previous efforts in modeling the main astrophysical/cosmological contributions to the WL, tSZ, and X-ray fields. Figure \ref{fig:Maps} shows an example of the maps produced by the above method. Any components that trace the density field are modeled using the simulated density field as a starting point, to ensure the correct correlations with other density-derived fields. Our WL modeling follows the sophisticated pipelines that have been applied to various lensing datasets \citep[\eg][]{Fluri2019DeepLearningKIDS, Gatti:2024:WPH, Cheng:2024:HSC_WST, Jeffrey:2025:Likelihood, Prat:2025:Homology}. The tSZ modeling broadly follows long-standing modeling pipelines in the community \citepalias[\eg][]{Omori2022Agora, Li:2022:WebskyRadio}, though forward-modeling approaches are yet to be applied to parameter inference with the tSZ.\footnote{Note, however, that CMB lensing efforts in these mm-wave surveys have successfuly performed fully forward-modeled parameter inference \citep{Ge:2025:SPTMUSE}.} The X-ray forward model we utilize is more simplistic than the other two, and thus requires additional investigation before being applied to data. Of particular note is the testing and/or incorporation of additional processing effects (masking, background subtraction, etc.) on map-level modeling. In the case of eROSITA, this is feasible as the associated processing pipeline\footnote{\url{erosita.mpe.mpg.de/edr/DataAnalysis}} is publicly available \citep[][see their Section 6]{Predehl:2020:eROSITA}.

\section{Primordial Signatures in the Nonlinear Universe}\label{sec:results}

\begin{table}
    \centering
    \begin{tabular}{c|c|m{10cm}}
    \hline
    Parameter & Fiducial Value & Explanation \\
    \hline
    $f_{\rm NL}$ & $0\pm 100, 0 \pm 10$ & The amplitude of PNGs. The size of the $\fNL$ changes according to the model to account for different normalizations\\[15pt]
    
    $\Omega_{\rm m}$ & $0.3175 \pm 0.01$ & The matter energy density fraction\\
    $\sigma_8$ & $0.834 \pm 0.015$ & Root-mean square of density fluctuations on the scale $R = 8 \mpc/h$\\[15pt]

    $A_{\rm IA}$ & $0.5 \pm 0.5$ & The amplitude of the NLA IA signal\\
    $\eta_{\rm IA}$ & $-0.1 \pm 0.1$ & The redshift scaling of the IA signal. \\[15pt]

    \hline
    
    $M_c$ & $10^{14 \pm 0.3}$ & Mass scale where the characteristic log-slope of the gas density profile is $\beta = -1.5$\\
    $\mu_\beta$ & $1 \pm 0.4$ & The mass scaling of the characteristic log slope of the gas density profile, $\beta$\\
    $\theta_{\rm ej}$ & $4 \pm 3$ & The radial scale of the ejected gas, $R_{\rm ej} = \theta_{\rm ej} \Rtwohc$\\[15pt]
    
    $\eta$ & $0.3 \pm 0.1$ & The power-law slope of the stellar-mass fraction with halo mass\\
    $\theta_{\rm co}$ & $0.1 \pm 0.05$ & The radial scale of the collapsed/core gas, $R_{\rm co} = \theta_{\rm co} \Rtwohc$\\
    $\gamma$ & $2.5 \pm 0.5$ & The GNFW slope for $r \sim R_{\rm ej}$\\
    $\delta$ & $7 \pm 3$ & The GNFW slope for $r \gg R_{\rm ej}$\\
    $\mu_{\theta_{\rm ej}}$ & $0 \pm 1$ & The power-law mass scaling of $\theta_{\rm ej}$\\
    $\nu_{\rm M_{\rm c}}$ & $0 \pm 0.1$ & The power-law redshift scaling of $M_c$\\
    $\nu_{\theta_{\rm ej}}$ & $0 \pm 0.1$ & The power-law redshift scaling of $\theta_{\rm ej}$\\[15pt]

    $\alpha_{\rm nt}$ & $0.2 \pm 0.1$ & The non-thermal pressure fraction at $\Rtwohc$\\[15pt]

    $Z_{\rm core}$ & $0.8 \pm 0.2$ & The metallicity in the halo core, in units of $\zsun$\\
    $Z_{\rm out}$ & $0.3 \pm 0.1$ & The metallicity in the halo outskirts, in units of $\zsun$\\[15pt]

    $f_{\rm CIB}$ & $1 \pm 0.5$ & A constant rescaling factor for the CIB field. Independent of frequency and scale.\\
    $f_{\rm AGN}$ & $1 \pm 0.5$ & A constant rescaling factor for the AGN field.\\[15pt]

    \hline
    
    \end{tabular}
    \caption{The parameters varied in our modeling. The broad classes of parameters, from top to bottom, are: (i) PNG, (ii) cosmology, (iii) IA, (iv) standard baryon SAM parameters, (v) extended baryon SAM parameters, (vi) Non-thermal pressure, (vii) Metallicity, and (viii) Foregrounds. For each, we show the fiducial value (which is used when estimating covariance), and the upper and lower shifts used when estimating derivatives. See \citetalias{Anbajagane:2024:MapBaryonification} for more details on the baryon SAM parameters.}
    \label{tab:parameters}
\end{table}

We now detail the signatures of primordial particle physics models in various observables of cosmic structure. In all cases, we utilize the Fisher information metric,
\begin{equation}\label{eqn:Fisher}
    \boldsymbol{F}_{ij} = \sum_{m,n}\frac{d\widetilde{X}_m}{d\theta_i}\big(\mathcal{C}^{-1}\big)_{mn}\frac{d\widetilde{X}_n}{d\theta_j},
\end{equation}
where $\frac{d\widetilde{X}_m}{d\theta_i}$ is the mean derivative of point $m$ in data vector $X$ with respect to parameter $\theta_i$. Then, $\mathcal{C}^{-1}$ is the inverse of the numerically estimated covariance matrix and includes the Hartlap correction factor \citep{Hartlap2007},
\begin{equation}\label{eqn:invertcov}
    \mathcal{C}^{-1} \rightarrow \frac{N_{\rm sims} - N_{\rm data} - 2}{N_{\rm sims} - 1} \,\mathcal{C}^{-1}.
\end{equation}
The Hartlap factor for all analyses in this work is $\gtrsim 0.9$. The datavector, $X$, consists of the moments of the field, computed on 10 scales ranging from $3.2\arcmin < \theta < 200\arcmin$. See Section \ref{sec:data:stats} for the precise definition of the statistics we use in this work.

The Fisher information defined in Equation \eqref{eqn:Fisher} can be artificially increased due to numerical noise in the derivative estimates, as this noise can incorrectly break parameter degeneracies; see \citet[][their Appendix A]{Coulton:2023:FisherBias} for a discussion of this. We check for this impact by varying the number of realizations used to compute the derivatives and the covariance. The Fisher information changes by $\lesssim 2\%$ ($\lesssim 10\%$) if we estimate the covariance (derivatives) using half the number of realizations. We discuss this further in Appendix \ref{appx:numconvergence}.

The parameters we vary are listed in Table \ref{tab:parameters}, alongside a brief description of their physical interpretation. The broader set of parameters in the \textsc{BaryonForge} modeling is presented in \citetalias{Anbajagane:2024:MapBaryonification}, alongside detailed discussions of the modeling choices and parameterizations. When an analysis does not vary any nuisance parameters, we will refer to it as fixed or ``fix.'' The combination of WL, tSZ, and X-ray is referred to as ``All.'' In all cases, we use the datavector defined in Section \ref{sec:data:stats} unless noted otherwise. The variant ``All++'' refers to the inclusion of the subset of fourth moments described in the same Section \ref{sec:data:stats}.

\subsection{Impact of nuisance parameters on WL}\label{sec:results:lensing}

\begin{figure}
    \centering
    \includegraphics[width=\columnwidth]{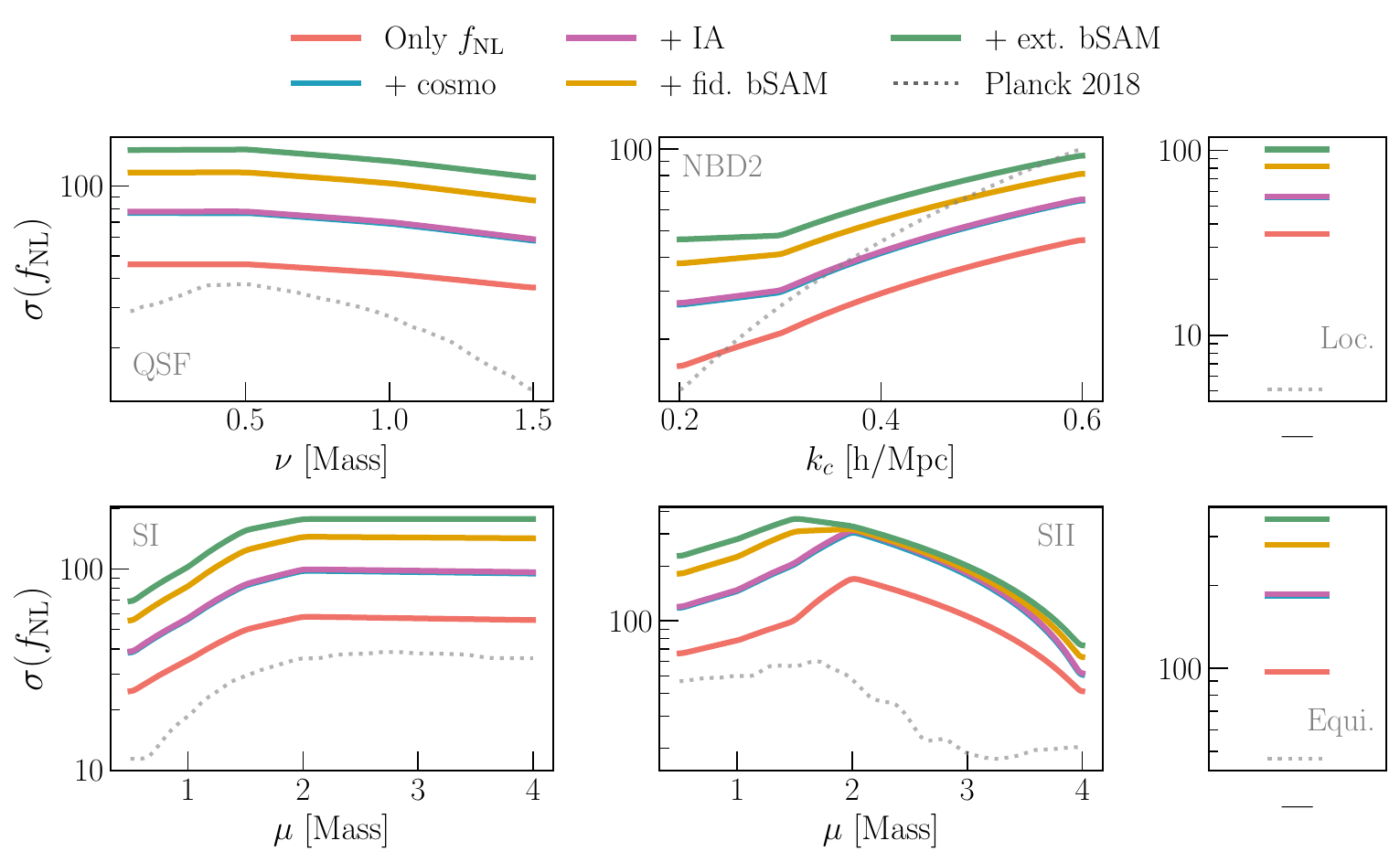}
    \caption{The Fisher information on $\fNL$ using the second and third moments of the lensing convergence from simulated LSST Year 10 data. We sequentially marginalize over broader sets of parameters, starting from cosmology, IA, a minimal baryon SAM, and an extended baryon SAM. See text for parameter set definitions. Marginalizing cosmology degrades constraints by about a factor about 70\%, marginalizing baryons parameters causes another 30\%, and marginalizing over all remaining parameters degrades it by 10\%. The total degradation is between factors of 2 to 3. The Planck constraints \citep{Planck2020PNGs, Sohn:2024:Colliders, paper2} are shown as dotted lines.}
    \label{fig:results:WL}
\end{figure}

We first generalize the results of \citetalias{paper1}, by incorporating a variety of nuisance parameters (IA and baryons) to our WL constraints on $\fNL$. \citetalias{paper1} were unable to do the same as the sparse set of simulations available for estimating derivatives meant multi-parameter analyses were less reliable. We have quadrupled the number of available simulations (from 30 to 120) for estimate the deriative with respect to $\fNL$, and thus can more reliably estimate degeneracies of $\fNL$ with nuisance parameters. The derivatives with respect to these nuisance parameters are estimated using 500 to 600 realizations.

Figure \ref{fig:results:WL} presents the impact of these parameter degeneracies on the WL-only constraints, and for a variety of different PNG models. The QSF, SI, and SII PNG models (see Section \ref{sec:data:models}) are parameterized by the particle mass, $\nu$ or $\mu$. The Local and Equilateral models have no free parameters. The red line shows a best-case scenario where all nuisance parameters are fixed; this represents the true statistical power of the LSST Y10 dataset, and is within 20\% to 50\% of the CMB constraints taken from \citet{Planck2020PNGs} and \citet{Sohn:2024:Colliders}. The WL constraints are particularly worse for the Local-type $\fNL$ as expected. This template has unique signatures on large scales but WL is predominantly sensitive to any PNG signal on smaller scales. Following \citetalias{Anbajagane2023Inflation}, WL is found to be more competitive for constraining Equilateral-type templates, and following \citet{paper2}, WL is a factor of 2 more constraining for the NBD models, where features get imprinted on small scales and are therefore probed more weakly by the CMB. We note that Figure \ref{fig:results:WL} compares only with the CMB, and that WL also provides competitive constraints when compared to other structure-based measurements such as those from galaxies \citep[\eg][]{Cabass2022MultifieldBOSS,Philcox2022BossPNG,Damico2022BossPNG}. See Figure 4 in \citetalias{Anbajagane2023Inflation} for a detailed comparison.

The other lines in Figure \ref{fig:results:WL} show the impact of marginalizing over other parameters. We sequentially add marginalization over cosmology ($\Omega_{\rm m}$ and $\sigma_8$), over IA ($A_{\rm IA}$ and $\eta_{\rm IA}$), over a fiducial baryon SAM set ($M_c$, $\theta_{\rm ej}$, $\mu_\beta$, $\alpha_{\rm nt}$, $Z_{\rm core}$, $Z_{\rm out}$, $f_{\rm AGN}$, $f_{\rm CIB}$), and the extended baryon SAM set which adds all remaining parameters in Table \ref{tab:parameters}. Note that the derivative of WL statistics with pressure, metallicity, and foregrounds parameters is zero as they only impact the thermodynamic fields. Marginalizing over cosmology has a noticeable impact, and removes roughly 40\% to 70\% of the available constraining power. Marginalizing over IA has almost no change relative to marginalizing over cosmology. The Year 6 cosmic shear analysis of DES also notes how fixing IA parameters only provides minor improvements to cosmology constraints \citep[][see their figure 7]{DES:2026:Y6Shear}, and we find similar behavior for $\fNL$. Marginalizing over baryons degrades the performance by 40\% to 50\%. Overall, the difference between the fixed analysis and fully marginalized analysis is a factor of 2.5.

\subsection{Synergies with tSZ and X-ray}\label{sec:results:multiwavelength}

We now extend the analysis to include the tSZ and X-ray fields. The WL field is sensitive to PNGs primarily via the latter's impact on the statistics of peaks in the density distribution \citep{Dalal2008ScaleDependentBias, Shirasaki2012fNL, Anbajagane2023Inflation}. A similar argument can be made for the tSZ and X-ray fields, which are dominated by cluster emissions \citep[\eg][]{Battaglia:2012:Gamma,McDonald:2026:XrayFlamingo}; hence our choice to include them in our analysis. In fact, the tSZ is particularly well-suited as its emission is dominated by cluster-scale halos even on large scales \citep[][see their Figure 7]{Battaglia:2012:Gamma}.

\begin{figure}
    \centering
    \includegraphics[width=\columnwidth]{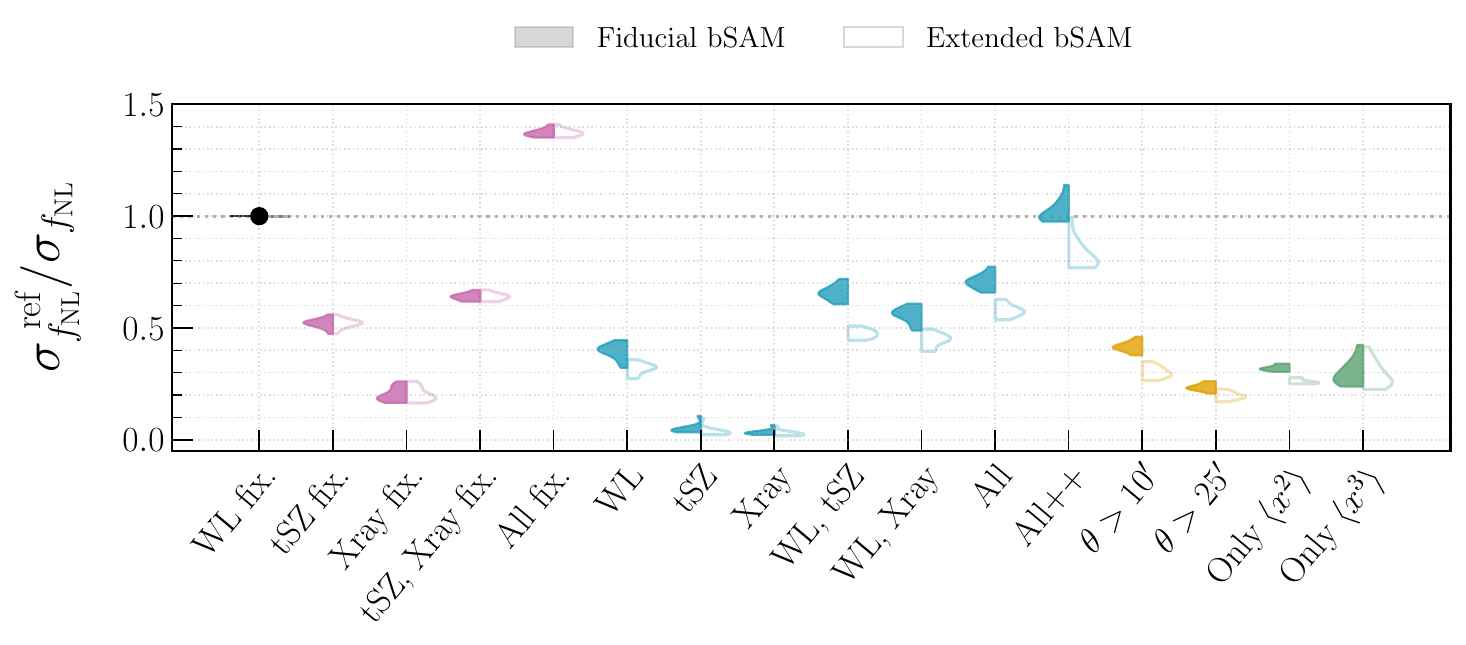}
    \caption{Relative improvement/degradation in constraints from various analysis configurations, relative to the fixed WL-only analysis which does not marginalize over any nuisance parameters. The violins show the distribution of ratios across the various models in Figure \ref{fig:results:WL}. The solid (open) violins are results from marginalizing over the fiducial (extended) baryon SAM. The purple (blue) points denote analyses with nuisance parameters fixed (varied). Here ``All'' refers to the second and third moments of the WL, tSZ, and X-ray fields, while ``All++'' denotes the inclusion of select fourth-order moments (Section \ref{sec:data:stats}) to the existing datavector. The yellow (green) points refer to changes to the scale cuts (summary statistics) relative to the ``All'' configuration. The solid/open violins marginalize over the fiducial/extended baryon SAM parameters (see Section \ref{sec:results:lensing} and Table \ref{tab:parameters}). The solid/open violins for the fixed analysis variants are the same as they do not vary any nuisance parameters in the first place. See Section \ref{sec:results:multiwavelength} and Section \ref{sec:results:ScaleNGs} for discussions of the results.}
    \label{fig:results:Variants}
\end{figure}

Figure \ref{fig:results:Variants} presents the improvement, relative to the fixed WL-only case, as we change the analysis configuration. We compute the relative change for all PNG models presents in Figure \ref{fig:results:WL} and show their distribution as violins. The open violins are relative improvements when marginalizing over only the fiducial baryon SAM rather than the extended one, as defined in Section \ref{sec:results:lensing}.

The purple violins in Figure \ref{fig:results:Variants} show the best-case constraining power from the tSZ and Xray fields (and combinations thereof). The tSZ-only case is half as precise as the WL-only one, while the X-ray-only case is an order of magnitude less precise. These behaviors are almost entirely set by the noise levels of each dataset. The tSZ is noise-dominated at all scales \citep[][see their Figure 4]{Raghunathan:2022:tSZnoise}, contrary to WL which is signal dominated on intermediate and larger scales. The eROSITA X-ray field is highly noise-dominated (even on larger scales, where the contribution from low-redshift massive clusters generates a shot-noise term; see \citet{LaPosta:2025:XraytSZWL}). The combination of tSZ and X-ray improves constraints significantly; the cross-correlation of the two fields provides more signal than the X-ray auto-correlation on its own. Under the best-case setup, the combination of WL, tSZ, and X-ray is 50\% better than the WL-only case. To mimic this increase using WL data we would need to double the size of the available dataset (\eg, from 14,000 $\deg^2$ to 28,000 $\deg^2$). Thus, there is clearly valuable information in these additional fields, sourced by the peaks-centric physical picture we discussed above.

The blue violins denote a more realistic case where we marginalize over all nuisance parameters. The inclusion of tSZ and X-ray data recovers constraining power lost in the WL-only analysis due to marginalizing over nuisance parameters. \citetalias{Anbajagane:2024:MapBaryonification} found a similar scenario in their LCDM analysis of LSST and SO; a WL plus tSZ analysis with nuisance-parameter marginalization delivered constraints similar to the WL-only analysis with all nuisance parameters fixed. Using only the tSZ or only X-ray, while marginalizing over cosmology and baryon SAM parameters, leads to no notable information on $\fNL$. Thus, the combination of probes is necessary to access the benefits of the tSZ and X-ray fields.

\begin{figure}
    \centering
    \includegraphics[width=\columnwidth]{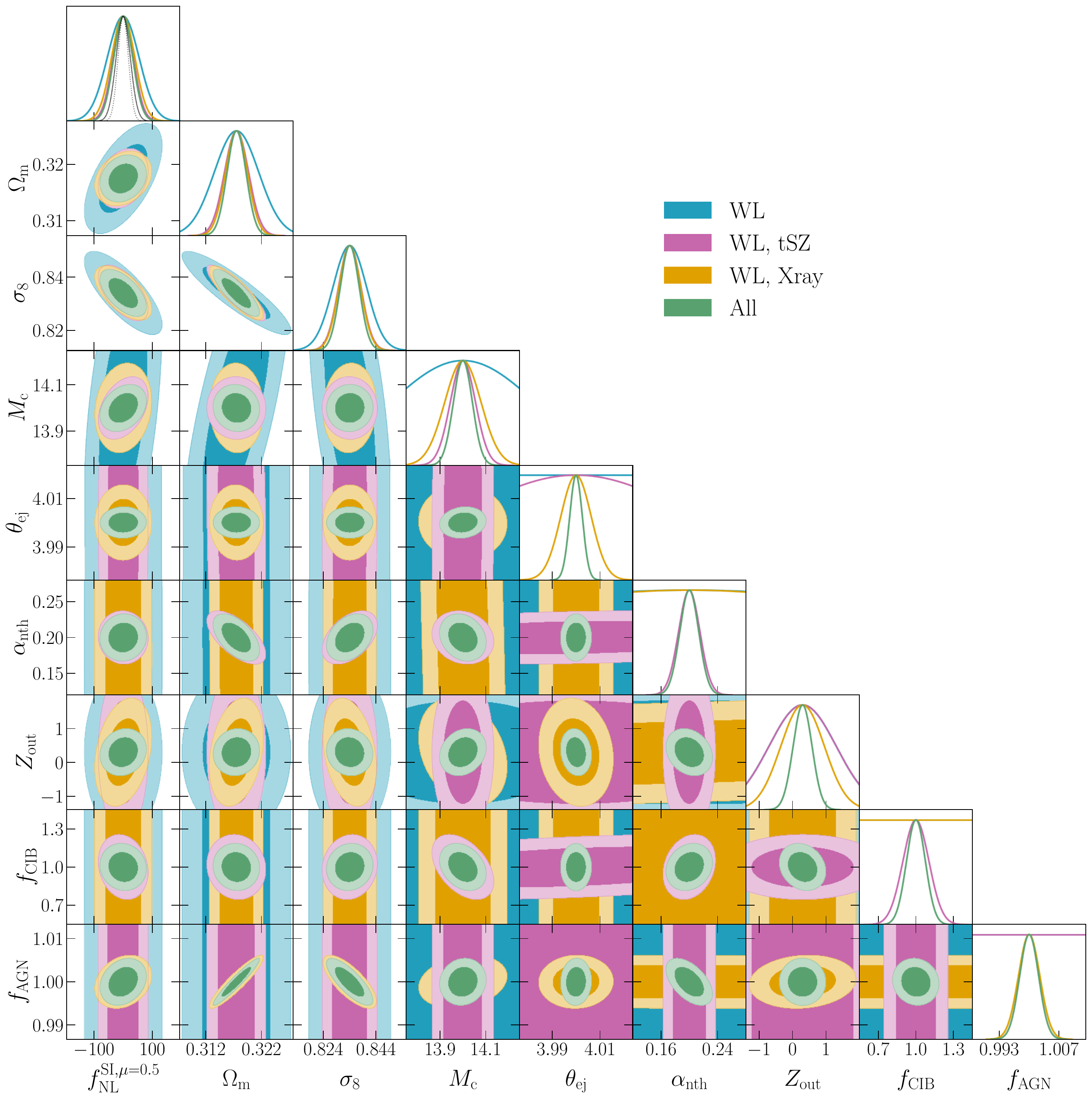}
    \caption{Fisher contours for various analysis setups. We also show 1D posteriors in an $\fNL$-only, ``fix.'' analysis from the ``WL only'' (black solid) and ``All'' (black dotted) data combinations. The joint analysis of different fields significantly breaks parameter degeneracies.}
    \label{fig:results:Contours}
\end{figure}

Figure \ref{fig:results:Contours} shows the parameter degeneracies, as determined by the Fisher information matrix, for various analyses setups. We only do so for an analysis varying $\fNLLoc$, but note that the discussion holds if we focus on other PNG models instead. We also only show a limited set of parameters for brevity sake, but stress that our analysis varies the full set described in Figure \ref{tab:parameters}. From our results, it is clear that the different fields (and their cross-correlations) are constraining different parameter combinations, so the combination of them all is far more constraining than would naively be expected. This follows from the LCDM analysis noted in \citetalias{Anbajagane:2024:MapBaryonification}.

We highlight four specific features from this triangle plot. First, the ejection scale of the gas profile, $\theta_{\rm ej}$, is determined at far better accuracy once both X-ray and tSZ are utilized. The probes on their own are sensitive to (different) combinations of gas density and temperature. Their combination breaks this degeneracy and places tighter constraints on the gas density, which is highly sensitive to the ejection radius.  Second, the constraints on $M_c$ show clear improvements as we add more fields. This highlights differences in the halo mass ranges that each probe is most sensitive to. The tSZ signal scales as $\Mtwohc ^ {5/3}$, whereas the X-ray signal scales like $\Mtwohc^{4/3}$ \citep{Kaiser:1986:Clusters}. The latter is flatter, and allows for more contribution from lower-mass halos.\footnote{The tSZ can also access information about lower-mass halos by pushing to higher-$\ell$ \citep[][see Figure 7]{Battaglia:2012:Gamma}. We are limited to $\ell < 2048$ in this work, and thus cannot fully extract this information.} 

Third, the non-thermal pressure is not well-constrained by the X-ray data. While it formally is sensitive to this term (due to how the temperature profile is modeled; see Equation \ref{eqn:temperature}), non-thermal pressure generally affects only the larger scales \citep{Nelson:2014:nonthermal} while the smaller scales are quickly thermalized due to the higher density (and therefore, shorter interaction timescales). X-ray maps are sensitive to emission from the halo core, and therefore are far less sensitive to non-thermal pressure compared to the tSZ. Instead, X-ray data are more sensitive to metallicty distributions. Fourth, we find the overall sensitivity to this halo metallicity is not large as the constraints are consistent with $Z_{\rm out} = 0$ within $1\sigma$. We have checked that our sensitivity is similarly weak for the other metallicity parameter we vary, $Z_{\rm core}$.

\subsection{Dependence on scale-cuts and non-Gaussian information}\label{sec:results:ScaleNGs}

Finally, we check the impact of scale cuts and non-Gaussian information on our constraints. This is presented as the yellow and green violins in Figure \ref{fig:results:Variants}. For reference, a scale of $\theta = 10\arcmin$ corresponds to comoving scales of $R = [2.5, 5.5, 10, 13]\mpc$ at $z = [0.3, 0.5, 1.0, 1.5]$, respectively. This is at, or significantly larger than, the scale setting the transition between the one-halo and two-halo regime, which is nominally the cluster virial radius \citep{Cooray2002HaloModel}. A cut of $\theta > 10\arcmin$ degrades constraints by 50\%, while $\theta > 25\arcmin$ does so by 70\%. Similar degradations were found in \citetalias{Anbajagane2023Inflation}. Note that the current lensing analyses reliably use scales down to $3 \arcmin$ in their $\xi_+$ datavector, which has the same $\theta$-to-$k$ dependence as a convergence field \citep[\eg][]{Secco2022Shear, Amon2022Shear, DECADE4, DECADE5, Wright:2025:KidsLegacy}. Inclusions of explicit baryon modeling, as we do here, then allows one to extend to all scales in the WL measurements \citep{Bigwood:2025:DESY3, DECADE5, Wright:2025:KidsLegacy}. Thus, the $\theta > 10\arcmin$ case discussed here is a highly conservative choice.

Next, we show the relative importance of Gaussian and non-Gaussian information in the field. Limiting the analysis to just the former (i.e. using only the second moments) removes about 60\% of our constraining power relative to the reference (``All''). Using just the third moments does not fare much better as parameter degeneracies deteriorate constraints on $\fNL$. However, the combination of these moments enables us to distinguish PNG signatures from those of cosmology and of the nuisance parameter models. As mentioned before, we have used only a small subset of the fourth moments (in the combination ``All++'') and are unable to test the utility of using all available fourth moments, as numerical convergence arise due to the significantly lengthened datavector. A larger set of simulations are needed to quantify this further. The benefits of non-Gaussian information are therefore understated in our analysis.

\subsection{Uncertainties from $\Lambda$CDM parameters}\label{sec:results:LCDMnuisance}

\begin{figure}
    \centering
    \includegraphics[width=\columnwidth]{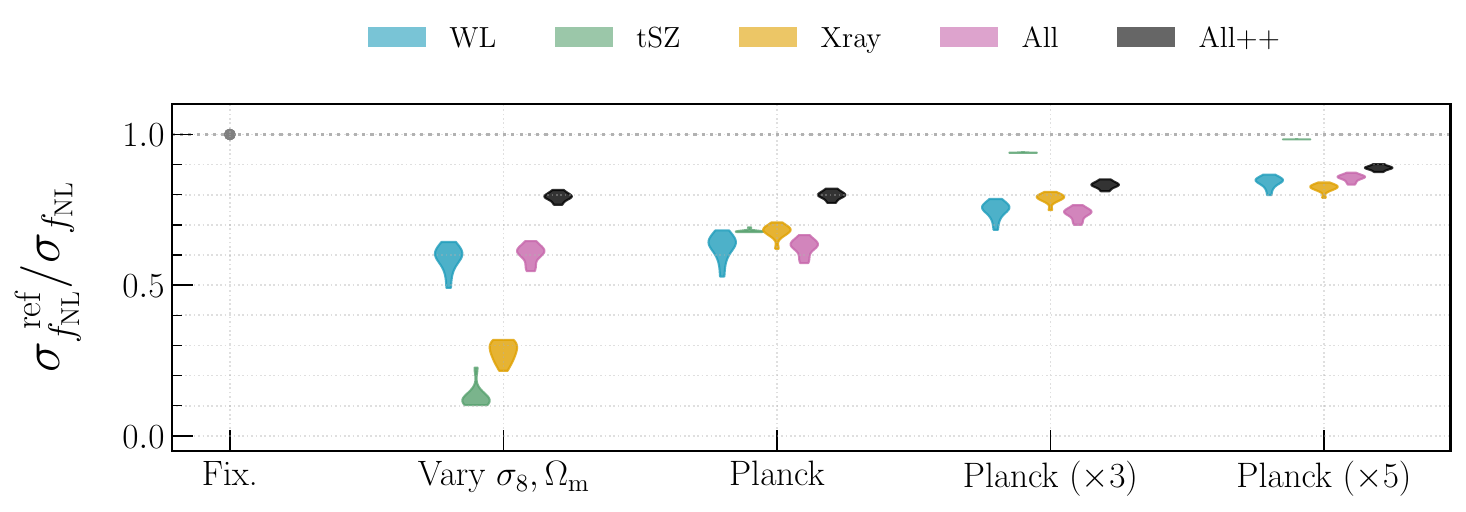}
    \caption{The Fisher information on $\fNL$ for various datavectors listed in Figure \ref{fig:results:Variants}, and the impact of marginalizing over $\sigma_8$ and $\Omega_{\rm m}$, under various priors. No other parameters are varied as we want to isolate the impact from the LCDM parameters alone. We use the ``Fix'' case as a baseline, and show the relative constraints from varying these two LCDM parameters with (i) uninformative priors, (ii) a \textit{Planck} prior from \citet{Planck:2020:CosmoParams}, (iii) a \textit{Planck} prior tightened by factor of 3, and a factor of 5.}
    \label{fig:results:cosmoMarg}
\end{figure}

Figure \ref{fig:results:cosmoMarg} details the impact of marginalizing over uncertainties in the LCDM parameters, $\sigma_8$ and $\Omega_{\rm m}$. Given these parameters control the growth of structure, they have degeneracies with $\fNL$. However, measurements of growth across scales are useful in this regard. As detailed in \citetalias{Anbajagane2023Inflation}, the impact of $\fNL$ on fields like WL is to alter the abundance of peaks observed in these fields. However, the statistics of the WL field on sufficiently large scales are insensitive to $\fNL$ and therefore can reliably calibrate LCDM parameters.\footnote{Under a given $\fNL$ model, the large-scale correlations of the density field will exhibit non-Gaussianities in the form of a bispectra. However, these are significantly suppressed in the lensing field as the latter is a projection of the density field over giga-parsec scales.} For this reason, the WL constraints degrade by only 40\% to 70\% (rather than a perfect degeneracy) when jointly constraining LCDM parameters and $\fNL$. In contrast, the tSZ and X-ray fields are dominated by cluster emission even on large scales and do not enjoy a similar self-calibration opportunity. Therefore, they suffer far stronger degeneracies between $\fNL$ and LCDM. Note that Figure \ref{fig:results:cosmoMarg} shows the relative change in constraining power \textit{within} each configuration and not that between different configurations.

Applying a prior of $\sigma(\sigma_8) = 0.007$ and $\sigma(\Omega_{\rm m}) = 0.005$, based on the final \textit{Planck} analysis \citep{Planck:2020:CosmoParams}, causes WL, tSZ, and X-ray to all degrade by only 30\%. Improving the precision of the prior by factors of 3 and 5 lowers the degradation to 20\% and 10\%, respectively. While these latter priors are unrealistic for a \textit{Planck}-only analysis, they can be achieved using the combination of high-precision data from other probes like supernovae, galaxy clustering, baryon acoustic oscillations, CMB lensing, etc. as well as from improvements to the CMB constraints facilitated by SO and SPT-3G \citep{Prabhu:2024:SPT, Abitbol:2025:SO}. It is also worth noting that such narrow priors still do not recover all the available information. The parameter $\fNL$ scales the density field like $\fNL \times \mathcal{O}(10^{-4})$, given it is second-order in the primordial potential field, whereas $\sigma_8$ and $\Omega_{\rm m}$ scale the field as $\mathcal{O}(1)$ as they are linear in this field. As a result, precise constraints on the LCDM parameters can still lead to some degradation (of around $10\%$) of $\fNL$ constraining power, relative to the case with infinite precision on the LCDM parameters.

\subsection{Future prospects}\label{sec:results:future}

Given our results above, we now discuss prospects for PNG inference using these surveys. We first list considerations that would improve our constraining power beyond what we have presented here:

\textbf{Larger sky coverage:} We have so far limited the tSZ and Xray data in our work to just the subset that overlaps with the LSST footprint. In the case of X-rays, eROSITA does not fully cover the LSST region. However, we can utilize the full-sky survey from the Rongten Satellite \citep[ROSAT,][]{Snowden:1997:ROSAT}, which was the precursor to eROSITA, and fill in this missing area. Furthermore, we can also use eROSITA and ROSAT data in the sky outside the LSST footprint to improve constraints from X-ray auto-correlations. A similar exercise can be performed for the tSZ by using data from the \textit{Planck} space mission \citep{Planck:2020:LegacyOverview}, which covers the full sky.\footnote{In all cases, we are implicitly assuming regions with strong Milky Way emission are removed from the survey area.} Even in the case of WL, any northern regions not covered by LSST can be filled in using existing data from DECADE \citep{DECADE5} and the Ultraviolet Near-Infrared Optical Northern Survey (UNIONS) lensing data \citep{Hervas-Peters:2026:Unions}, or upcoming data releases from the \textit{Euclid} space mission \citep{Euclid}, with the latter expected to have the highest statistical power of the three.

\textbf{Pushing to smaller scales:} All our maps have been limited to NSIDE = 1024, whereas we could access smaller scales by increasing it to NSIDE = 2048 or 4096. This will be particularly important for the tSZ, whose signal peaks at $\ell \approx 3000$, but will also be relevant for WL where the observational systematics are still well under control on such scales \citep[\eg][]{y3-shapecatalog, Yamamoto2025, DECADE1}. Lensing measurements on such scales are routinely used for inference \citep{Wright:2025:KidsLegacy, DECADE5, Bigwood:2025:DESY3}. In this work, we are limited in exploring this option as our density shells are all saved at NSIDE = 1024.
    
\textbf{Improving simulation resolution:} As discussed before, the lower-resolution nature of our simulation underestimates the signal in the different fields. It does this (i) by effectively smoothing the field, thereby suppressing the matter power spectrum on smaller scales and reducing our WL signal; and (ii) by limiting our halo catalogs to $\Mtwohc > 10^{14} \msun/h$, which then misses the 10-20\% of the signal contributed by group-scale objects for the X-ray field \citep[][see their Figure 3]{McDonald:2026:XrayFlamingo} and 10\% of the signal they contribute to the tSZ field \citep[][see their Figure 7]{Battaglia:2012:Gamma}. Improving simulation resolution will overall increase our signal.

\textbf{Utilizing more complete summary statistics:} In this work, we have used a simpler set of summary statistics, the moments, motivated by their simple calculation and their straightforward connection to analytic models. However, recent analyses show the Figure of Metric --- defined as the inverse of the area spanned by the 2D marginal posterior --- of $\sigma_8$ and $\Omega_m$ can be increased by 60\% to 80\% using statistics beyond just the moments \citep{Gatti:2024:LFIResults, Jeffrey:2024:LFIResult}. We expect such improvements to translate to constraints on $\fNL$ as well given this parameter, like $\sigma_8$ and $\Omega_m$, has a dominant impact on the growth of structure.

Now, we move to discussing potential concerns in our modeling approach. While our model is fairly sophisticated, we have made a few simplifying assumptions in our approach. We list these now, and discuss why we do not expect them to have a significant impact on our narrative in this work.

\textbf{Foreground model flexibility:} Our models for the AGN, Radio, and CIB foregrounds assume some prescription from the literature (e.g., transfer functions, luminosity functions, etc.) and apply that to our simulations. We have parameterized these foregrounds by a free amplitude, $f_i$, and marginalize over them in our analysis. While this adds significant flexibility, one could also vary the parameters defined in these models from the literature, rather than treating them as fixed. This could, in principle, notably alter the scale-dependence or frequency-dependence (in the case of radio and CIB foregrounds) of the signal. In practice, we find that marginalizing over $f_{\rm CIB}$ broadens $\fNL$ constraints by 10\%. We therefore expect that additional flexibility in the CIB models will degrade constraints by $\mathcal{O}(10\%)$. Though in practice, the effect may be smaller since the nuisance parameters have stronger degeneracies with each other than they do with $\fNL$. Marginalizing over $f_{\rm AGN}$ has negligible impact as it is well-constrained by the cross-correlations (Figure \ref{fig:results:Contours}).

\textbf{Baryon model flexibility:} Our extended baryon SAM parameter set is more flexible than models applied to data \citep[\eg][]{Pandey:2025:ACTDES, DECADE5, Bigwood:2025:DESY3, Dalal:2026:Baryons}, but it is still possible that one needs to vary other parameters in the model that are currently being held fixed. Given we are already varying a wide set of astrophysical quantities (stellar fraction relations, gas profile slopes, etc.), we do not think this addition will degrade $\fNL$ constraints appreciably. In Figure \ref{fig:results:WL}, the difference in constraints under the fiducial and extended baryon SAM models (which have eight and fifteen parameters, respectively) are at or below the 10\% level so we expect additional extensions to have negligible impact. Instead, we anticipate the degeneracies between the various baryon SAM parameters will broaden as the new parameters will necessarily control the same underlying gas/star distributions as the existing parameters.

\textbf{Intrinsic Alignments:} Throughout this work we have utilized the NLA model of IA. This is the leading-order prediction for galaxy IA, and can subsequently be extended by incorporating additional, higher-order terms representing various facets of galaxy formation \citep{Blazek2019TATT, Chen:2024:Spinosaurus}. In this work, we have been limited to the NLA model as the other models require simulated quantities beyond the density fields available to us here. While the community has documented the impact of extended IA models on cosmology constraints from two-point WL statistics, the same on constraints obtained from higher-order WL statistics is far less known. \citet{Gomes:2026:TATT} have shown that the joint analysis of two-point and three-point functions can significantly break degeneracies in the extra parameters introduced in these extended models (relative to NLA). In two-point WL statistics, the Figure of Merit in $\sigma_8$ and $\Omega_{\rm m}$ drops by 25\% when going from the simpler NLA model to the more complicated TATT model \citep[][see their Table III]{DES:2026:Y6Shear}. Based on \citet{Gomes:2026:TATT}, constraints from higher-order WL statistics will degrade by less than 25\% when using the extended IA models. In our analysis, we find the inclusion of tSZ and Xray improves constraints on $A_{\rm IA}$ by 15\%, indicating a potential to self-calibrate IA nuisance parameters using multi-wavelength analyses. Overall, our discussion here is limited by current understanding of IA prescriptions for simulation-based modeling. However, using our current analyses and results, we expect that switching from NLA to an extended IA model --- for a multi-wavelength analysis using higher-order statistics --- can degrade $\fNL$ constraints by $\lesssim 10\%$.

\textbf{In summary}, given the above discussion, the prospects for multi-wavelength analyses of PNGs are optimistic. It is also worth noting that we have not yet considered two other key, existing probes of the cosmic baryon distribution --- the kinematic SZ effect \citep{Carlstrom2002SZReview}, and fast radio bursts \citep[FRBs,][]{Petroff:2022:FRBs}. The former has already placed informative constraints on the distribution of baryons outside halos \citep{Bigwood:2024:BaryonsWLkSZ, Siegel:2025:Baryons}, and the latter are expected to be complementary to established methods for constraining the baryon evolution \citep{Reischke:2023:FRBs, Wayland:2026:FRB3x2pt, Sharma:2026:FRB3x2pt}. Including these probes into the forward-model is not immediately possible as they require the gas density field as an input, and their signal is not dominated by massive clusters the same way the tSZ and X-ray signals are, i.e. the diffuse gas component is critical but cannot be modeled by painting profiles around massive halos. While it is possible to use baryon SAM models to generate gas fields \citep{Schneider:2025:Baryonification}, including contributions to the diffuse component, the accuracy of this prescription is yet to be determined.

\section{Conclusions}\label{sec:conclusions}

PNGs encompass a wide variety of signatures in the primordial density field, arising from an equally wide variety of particle physics processes in the early Universe. Much has already been uncovered about which signatures, and their respective amplitudes, are consistent with our current observations of the Universe. This has been chiefly determined using observations of the CMB, and more recently using spatial correlations in the positions of galaxies.

In this work, we systematically explore the PNG information present in the moments of the WL convergence, tSZ, and X-ray photon count maps. We utilize a simulation-based approach with a flexible baryon SAM, and explicitly model the relevant tSZ and X-ray foregrounds. Our approach for the latter uses the simulated density field as a starting point, ensuring the foregrounds are all correlated with the cosmic signal and with each other. We then use the moments of the field to summarize these observations, and quantify the constraining power on $\fNL$ and various other parameters.

\noindent We summarize our findings as follows:
\begin{itemize}
    \item The raw statistical power of the WL convergence maps from LSST Year 10 are sufficient to deliver PNG constraints comparable to current \textit{Planck} constraints, and a factor of two better than \textit{Planck} for models with signals peaking on smaller scales. However, marginalizing over cosmology, IA, and Baryons leads to degradation by a factor of 2 to 3 depending on the model (Figure \ref{fig:results:WL}).

    \item The tSZ and X-ray fields --- via their sensitivity to the formation and evolution of massive halos --- contain information on PNGs, but this is largely inaccessible once we also vary nuisance parameters in the astrophysics models (Figure \ref{fig:results:Variants}). Instead, the combination/cross-correlation of fields is required for accessing $\fNL$ information.

    \item A joint analysis of WL, tSZ, and X-ray --- while marginalizing over all nuisance parameters (Table \ref{tab:parameters}) --- can still be more precise than a no-nuisance, $\fNL$-only analysis using WL alone (Figure \ref{fig:results:Variants}). Thus, a multi-probe analysis can self-calibrate the nuisance models while also improving the raw constraining power on $\fNL$.

    \item The inclusion of tSZ and X-ray provides significant degeneracy breaking, primarily in the baryon SAM parameters, which naturally has downstream impacts on $\fNL$ constraints from WL (Figure \ref{fig:results:Contours}).
    
\end{itemize}

The prospects for multi-wavelength analyses of PNGs are optimistic, and the results of this work highlight the relevance and synergy of such analyses. For the reasons discussed in Section \ref{sec:results:future}, the above forecast is still an underestimate of the total statistical power available in the current/upcoming data. Utilizing simulation-based inference will allow us to better employ these multi-wavelength data in constraining the physics of the early Universe.

More broadly, this work highlights a growing opportunity to elevate PNG constraints into the same multi-probe stature enjoyed by LCDM analyses, where $\mathcal{O}(10)$ probes --- together spanning a wide range in redshift, mass, and length scale --- are used to set constraints on the early Universe. Existing probes of PNGs (CMB and galaxy clustering) are limited to scales of $k \lesssim 0.1 \mpc/h$, but there remain well-motivated models of the early Universe that impart signatures on scales well above this limit \citepalias[\eg][]{paper1, paper2}. In this work, the NBD2 model is one such example. Using other established probes of LCDM (such as WL) to constrain PNGs will extend our scale sensitivity by at least one order of magnitude, with the potential to extend it even further as we include other probes like dwarf galaxies and strong lensing. Now is a particularly opportune time to pursue such efforts, given the wealth of (multi-wavelength) data that will be available in the next decade \citep{Spergel:2015:Roman, LSST2018SRD, Euclid}

The physics of the early Universe is rich in its phenomenology and is foundational to our understanding of the origin and evolution of the Universe. The cosmology community has spent decades on shaping the modern landscape of multi-probe data analysis. Tackling primordial questions with this same infrastructure will provide a significant leap in our access to information, and hopefully, in our understanding of the early Universe.

\section*{Acknowledgements}

This paper, submitted as a single-author work, satisfies the thesis requirement at the University of Chicago. DA thanks a large community of colleagues, mentors, and friends at UChicago, as well as at UMichigan and Penn, for their guidance and collaborations throughout the projects that eventually led to this program. DA is immensely grateful to Chihway Chang, for her mentorship, her friendship, and for giving him an ideal to aspire toward. He thanks Marco Gatti, Bhuv Jain, Shivam Pandey, Yuuki Omori, Hayden Lee, Austin Joyce, Alex Drlica-Wagner, Josh Frieman, Matt Becker, Robert Gruendl and innumerable others for their guidance and collaboration throughout his graduate career. He also regrets to inform his eternal ping-pong-fanatic colleague that he has not ``solved cosmology'' as the latter once joked, but will continue to make his best effort :)

Finally, DA acknowledges the tales of the rabbit \& the monkey, the stagnant stock broker, and the all-seeing I.M.E. protocol, which were all instrumental in the completion of this work and of this PhD.

DA was supported by the National Science Foundation
(NSF) Grant No. AST-2508321, and is grateful to the NSF for prior support during his PhD from the NSF Graduate Research Fellowship under Grant No. DGE
1746045 and from NSF Grant No. AST-2108168. All analysis in this work was enabled greatly by the following software: \textsc{Pandas} \citep{Mckinney2011pandas}, \textsc{NumPy} \citep{vanderWalt2011Numpy}, \textsc{SciPy} \citep{Virtanen2020Scipy}, and \textsc{Matplotlib} \citep{Hunter2007Matplotlib}. We have also used
the Astrophysics Data Service (\href{https://ui.adsabs.harvard.edu/}{ADS}) and \href{https://arxiv.org/}{\texttt{arXiv}} preprint repository extensively during this project and the writing of the paper.

\section*{Data Availability}

All pipelines and datasets used in this work are publicly available. 
\begin{itemize}
    \item The \href{https://ulagam-simulations.readthedocs.io}{\textsc{Ulagam}} suite contains all simulations presented and utilized in this work: {ulagam-simulations.readthedocs.io}. 
    \item \href{https://github.com/DhayaaAnbajagane/Aarambam}{\textsc{Aarambam}} provides all methods used to construct initial conditions for the PNG simulations: {github.com/DhayaaAnbajagane/Aarambam}.
    \item \href{https://github.com/DhayaaAnbajagane/BaryonForge}{\textsc{BaryonForge}} is the baryon semi-analytic model that consistently models WL, tSZ, X-ray, etc. observables both via halo-model approaches and via simulation forward models: {github.com/DhayaaAnbajagane/BaryonForge}
    \item \href{https://github.com/DhayaaAnbajagane/Vaanam}{\textsc{Vaanam}} converts simulations into synthetic skies for  WL, mm-wave, and X-ray surveys: {github.com/DhayaaAnbajagane/Vaanam}
\end{itemize}

\noindent Please contact DA if you are interested in other products from the simulations, or need assistance using one of the above codebases.

\bibliographystyle{mnras}
\bibliography{References}

%%%%%%%%%%%%%%%%%%%%%%%%%%%%%%%%%%%%%%%%%%%%%%%%%%

%%%%%%%%%%%%%%%%% APPENDICES %%%%%%%%%%%%%%%%%%%%%
\newpage
\appendix

\section{Additional modeling details}\label{appx:model}

We now provide a more detailed description of the modeling pipeline from Section \ref{sec:mockmaps:mocks}. We discuss the WL, tSZ, and X-ray observables separately in Sections \ref{appx:model:WL}, \ref{appx:model:tSZ}, and \ref{appx:model:xray}, respectively.

\subsection{WL modelling}\label{appx:model:WL}

As noted before, WL is the most mature of the three observables considered in this work when it comes to simulation-based inference on data. Many groups have independently produced simulations and WL modeling pipelines that have accurately matched the statistics of noisy datasets from lensing surveys \citep[\eg][]{Fluri2019DeepLearningKIDS, Fluri2022wCDMKIDS,Kacprzak2023Cosmogrid, Gatti:2024:LFIValidation, Cheng:2024:HSC_WST, Thomsen:2026:SBI}. As a result, there is a well-established and well-tested procedure for forward modeling the WL field.

\textbf{The lensing convergence field, $\boldsymbol{\kappa}$}, can be constructed directly from the density shells as,
\begin{equation}\label{eqn:convergence_definition}
    \kappa(\nhat, z_s) = \frac{3}{2}\frac{H_0^2\Omega_{\rm m}}{c^2}\int_0^{z_s}\!\!\!\delta_{\rm bcm}(\nhat, z_j) \frac{\chi_j(\chi_s - \chi_j)}{a(z_j)\chi_s}dz_j\frac{d\chi}{dz}\bigg|_{z_j},
\end{equation}
where $z_s$ is the redshift of the ``source'' plane/galaxies being lensed, $\nhat$ is the pointing direction on the sky, $\chi$ is the comoving distance from an observer to a given redshift, $a$ is the scale factor, $H_0$ is the Hubble constant, $\Omega_{\rm m}$ is the matter energy density fraction at $z = 0$, and $c$ is the speed of light. Here, we have used the shorthand $\chi(z_s) \equiv \chi_s$ and $\chi(z_j) \equiv \chi_j$. Note that $\deltabcm$ is the density field with baryon corrections, modeled using the baryon SAM described in Section \ref{sec:mockmaps:SAMs}.

\textbf{Intrinsic alignments} characterize the spatial correlations of galaxy shapes in the absence of any lensing effects, and must therefore be accounted for \citep{Troxel2015IAReview, Lamman2023}. Such correlations are induced, at linear order, by the tidal field. We include this contribution as
\begin{equation}\label{eqn:IA}
    \kappa_{\rm IA}(\nhat, z_s) = -\frac{A_{\rm IA} \rho_{\rm crit, 0} \Omega_{\rm m}}{D_+(z_s, \Omega_{\rm m})}  \bigg(\frac{1  + z_s}{1 + 0.6}\bigg)^{\eta_{\rm IA}} \delta_{\rm bcm}(\nhat, z_s),
\end{equation}
where $\kappa_{\rm IA}(\nhat, z_s)$ is the convergence signal corresponding to IA, $A_{\rm IA}$ is the amplitude of the IA effect, $\rho_{\rm crit}$ is the critical density of the universe at $z = 0$, $D_+$ is the linear growth function normalized to $D_+(z = 0) = 1$, and $\eta_{\rm IA}$ is the redshift scaling. Both $A_{\rm IA}$ and $\eta_{\rm IA}$ are free parameters of the IA model. The IA signal can trivially be added to the cosmology signal as $\kappa \rightarrow \kappa + \kappa_{\rm IA}$. This IA parameterization is called the Non-linear Linear Alignment (NLA) model\footnote{This somewhat oxymoronic naming is because the model is the ``linear'', or first-order, IA correction but uses the ``non-linear'' density field}. Other, more sophisticated parameterizations of the IA effect also exist \citep{Blazek2019TATT, Chen:2024:Spinosaurus}, but are currently not straightforward to simulate via the approach above as they require additional fields beyond just the projected density shells available in this work.\footnote{\citet{HarnoidDeraps:2025:TATTsims} provide an avenue for utilizing density shells to approximately simulate the Tidal-Torquing Tidal Alignment (TATT) model, though they find a few limitations in doing so using projected shells. For this work, we stick to the simpler NLA model.} In Section \ref{sec:results:future}, we discuss some arguments as to why our constraints are fairly insensitive to changes in these modeling choices.

\textbf{The lensing shear}, and not the convergence, is the actual observable in a WL survey. Such a survey measures galaxy shapes, which primarily trace the shear field, $\gamma$, and not the convergence field, $\kappa$. However, the shear and convergence field can be transformed into each other using the Kaiser-Squires (KS) transform \citep{Kaiser1993KS}, implemented in harmonic space as
\begin{equation}\label{eqn:Kappa2Shear}
    \gamma^{\ell m}_E + i\gamma^{\ell m}_B = -\sqrt{\frac{(\ell + 2)(\ell - 1)}{\ell(\ell + 1)}} \bigg(\kappa^{\ell m}_E + i\kappa^{\ell m}_B \bigg),
\end{equation}
where $X_{\{E, B\}}$ are the E-mode and B-mode (or Q and U polarizations, in healpix notation) of the field. However, in actuality, galaxy shapes trace the \textit{reduced shear}, $\gamma \rightarrow \gamma / (1 - \kappa)$. We include this transformation in our modeling. We also explicitly include the magnification effect via $\gamma \rightarrow \gamma (1 + q\kappa)$, which accounts for the (observed) source galaxy density fluctuations being correlated with the foreground lensing due to magnification effects. Here, $q$ is the magnification bias, which varies according to the source sample. We fix $q = 2$, which is the median value exhibited in the source galaxy sample from DES Year 6 \citep[][see their Table IV]{Legnani:2026:Y6Magnification} when evaluated on the DES synthetic source injection dataset, which forward models the survey images with the relevant systematics \citep{Anbajagane:2025:Y6Balrog}.

\textbf{Shape noise} is the dominant contribution to a lensing map on smaller scales. Upon generating the two shear fields, $\gamma^{1, 2}$, we add shape noise in real space. The forward-modeled field includes Gaussian shape noise with a standard deviation of
\begin{equation}\label{eqn:shapenoise}
    \sigma_\gamma = \frac{\sigma_e}{\sqrt{n_{\rm gal}A_{\rm pix}}},
\end{equation}
where $n_{\rm gal}$ is the source galaxy number density, and $A_{\rm pix}$ is the pixel area for a given map resolution. The per-galaxy shape noise is taken to be $\sigma_e = 0.26$. We then simply draw a random variate per pixel, $\gamma^{1, 2}_{\rm noise} \sim \mathcal{N}(0, \sigma_e^2)$.

\textbf{Source clustering} is the final higher-order shear effect we consider in this work. Source galaxies exhibit cosmological clustering\footnote{We distinguish this from the apparent clustering of source galaxies due to the angular variations in survey observing conditions (\eg depth, seeing, etc.) across the sky. In simulation-based analyses of DES data, such effects are included by fixing the positions of source galaxies in the forward model to those of the actual dataset \citep[\eg][]{Gatti2022MomentsDESY3}. A more principled approach can use synthetic source injection products from these surveys \citep{Everett:2020:BalrogY3, Anbajagane:2025:Y6Balrog} to build the selection function of source galaxies and include it explicitly in the forward model.} that is correlated with the same density fields that generate the lensing signal. This is accounted for following the prescription of \citet{Gatti2023SC},
\begin{equation}\label{eqn:SC}
    \gamma^{1, 2}_{\rm SC}(\nhat) = \frac{\int n(z) (1 + b_g \deltabcm(\nhat, z)) \gamma^{1, 2}(\nhat, z)dz}{\int n(z) (1 + b_g \deltabcm(\nhat, z))dz},
\end{equation}
where $n(z)$ is the source redshift distribution of the tomographic bin, averaged across the survey footprint, $\delta(\nhat, z)$ and $\gamma(\nhat, z)$ are the density and true shear maps at a given direction/pixel and redshift, and $b_g$ is the source galaxy bias. In simple terms, equation \eqref{eqn:SC} modulates the $n(z)$ across the survey footprint by reweighting it in a direction-dependent way using the density fields. We assume $b_g = 1$ following \citet{Gatti2023SC} since source galaxies are on average blue galaxies \citep[\eg][]{McCullough:2024:Blue} and therefore not highly clustered. Equation \eqref{eqn:SC} focuses on the cosmological signal, whereas the noise field must also be corrected as the directional modulation alters the number of source galaxies in a pixel. This changes the noise variance of the inferred shear field in a given sightline. We implement this correction as,
\begin{equation}\label{eqn:SCNoise}
    \gamma^{1, 2}_{\rm SC,\, noise}(\nhat) = \gamma^{1, 2}_{\rm noise}(\nhat) \sqrt{\frac{\int n(z)dz}{\int n(z) (1 + b_g \delta(\nhat, z))dz}} ,
\end{equation}

Finally, we apply the survey mask to the observed, noisy shear fields, and inverse-transform back to the convergence using the inverse of Equation \eqref{eqn:Kappa2Shear}. This final map contains all relevant mask and noise effects that are present when converting shear fields to convergence. Contributions from other factors --- such as contamination from the point-spread function (PSF), B-mode residuals, etc. --- have been extensively tested \citep[\eg][]{Gatti2020Moments, Anbajagane2023CDFs} and are not necessary.\footnote{This statement is predicated on such effects (PSF contamination, B-mode contributions, etc.) being sufficiently small for the standard two-point correlation analyses, as determined by tests used in current surveys \citep[\eg][]{y3-shapecatalog, Zhang:2023:HSCPSF, DECADE1, DECADE3, Wright:2025:KidsLegacy}.}

\subsection{Thermal SZ modelling}\label{appx:model:tSZ}

The thermal SZ effect is the cosmic distribution of thermal electron pressure integrated along the line-of-sight. Under the assumption of hydrostatic equilibrium, thermal gas pressure can be predicted using the gas density distribution and the total matter density distribution. Both these components are already specified in the baryon SAM described in Section \ref{sec:mockmaps:SAMs}, and therefore are readily available for use in predicting the thermal pressure. \citetalias{Anbajagane:2024:MapBaryonification} provide an implementation that self-consistently predicts thermal gas pressure, and generates full-sky maps of the tSZ. We also know that all objects are not in perfect hydrodynamical equilibrium, and have some non-thermal contributions, such as turbulence \citep[\eg][]{Nelson:2014:nonthermal}. We model this flexibly using a non-thermal profile prescription. 

Our main expression for thermal gas pressure is now given as,
\begin{equation} \label{eqn:dPdr_HE}
    P_{\rm gas, th}(R) = \bigg[1 - \alpha_{\rm nt} f_z \bigg(\frac{R}{\Rtwohc}\bigg)^{\gamma_{\rm nt}}\bigg]\int^{R}_{\infty}dr\frac{GM^{\rm 1h}_{\rm tot}(<r)}{r^2}\rho_{\rm gas}(r)
\end{equation}
where the integral solves for the total pressure required to support the gas at a certain halo-centric radius, given the enclosed mass distribution of the 1-halo component (i.e. ignoring large-scale features from neighboring halos). The term in the square brackets removes the non-thermal component. The latter model is taken from \citet{Shaw:2010:Nonthermal}, and the parameters are defined as $f_z = \min[(1 + z)^{\nu_{\rm nt}}, (f_{\rm max} - 1)\tanh(\nu_{\rm nt}z) + 1],$ and $f_{\rm max} = 6^{-\gamma_{\rm nt}}/\alpha_{\rm nt}$, where $\alpha_{\rm nt}$ is the amplitude of non-thermal pressure, $\gamma_{\rm nt}$ its scaling with radius, and $\nu_{\rm nt}$ its scaling with redshift. We set $\gamma_{\rm nt} = 0.5$ and $\nu_{\rm nt} = 0.3$ following \cite{Pandey:2025:Godmax}. The above integral assumes the boundary condition $P_{\rm gas,th}(R = \infty) = 0$.

We convert the gas pressure to electron pressure using the cosmic hydrogen and helium abundances, $P_{\rm e, th}(r) = \frac{4 -2Y}{8 - 5Y}P_{\rm gas, th}(r)$ where $Y = 0.24$ is the cosmic helium abundance.\footnote{While we explicitly include metallicity effects in the X-ray modeling discussed in Appendix \ref{appx:model:xray}, it can be safely ignored here as it provides a $<1\%$ correction to this transformation.} This conversion assumes the temperature of the electron and protons (or gas) are in equilibrium. Recent work in tSZ data has shown hints of non-equilibrium at the cluster virial radius that is consistent with features arising from accretion shocks \citep{Anbajagane2022Shocks, Anbajagane2023Shocks} and also consistent with behavior found in hydrodynamical simulations \citep{Rudd2009ElectronNE, Avestruz2015ElectronNE}. Other works in X-ray, radio, gamma-ray, etc. observations have found different features at similar radii \citep[\eg][]{Hurier2019ShocksSZPlanck, Zhu2021ShockXray, Hou2023ShockRadio}, though these works do not consider/discuss a link to temperature non-equilibrium. In \citetalias{Anbajagane:2024:MapBaryonification} (see their Figure 3), the shock features in the gas density --- which generically induce a $4\times$ dropoff in the density beyond the shock radius, as set by the Rankine-Hugonoit jump conditions \citep{Rankine1870, Hugoniot1887} for a high-mach number shock in a mono-atomic gas --- are shown to impact the baryon SAM-based corrections by up to a factor of 2. Since much remains uncertain about the nature of shocks in these outskirts, our current modeling does not include it. Future works will need to check that such features do not bias inference, as shocks will lead to correlated features across tSZ and WL.\footnote{We do not expect shock features to be relevant for X-ray modeling as this feature is in the cluster outskirts where data is generally noise dominated except for cases where individual, nearby systems are observed with megasecond exposure times. So this shock feature is less relevant for wide-field X-ray data than in the tSZ case.}

Using our model for $P_{\rm e, th}$, we can obtain the thermal SZ through a projection integral along the line-of-sight,
\begin{equation} \label{eqn:comptonY}
    P_y(r, z) = \frac{\sigma_T}{m_e c^2}\int_{-\infty}^{\infty}\frac{d\chi}{1 + z}P_{\rm e, th}\bigg(\sqrt{\chi^2 + r^2}\bigg),
\end{equation}
where $\sigma_T$ is the Thomson scattering cross-section, and $m_ec^2$ is the rest mass energy of the electron. The profile $P_{\rm e, th}$ is in physical units. Thus, given a halo lightcone catalog and the above profiles, we can generate a sky map of the thermal SZ effect. See \citetalias{Anbajagane:2024:MapBaryonification} for more details on the algorithm for this, which includes pixel window functions, shell widths, etc.

The actual raw observable in a mm-wave survey is the measured sky brightness at various frequency bands. These observations contain contributions from a wide variety of emissions, not just the tSZ, and many of the contributions are also correlated with the cosmic density field (and therefore, the lensing field). The key foregrounds in the mm-sky are the CMB primaries, the kinematic SZ effect, the Cosmic Infrared Background (CIB), radio sources, and the thermal/atmosphere noise. All of the above were included in the analysis of \citetalias{Anbajagane:2024:MapBaryonification} using a data-constrained analytic model \citep{George:2015:foregrounds}. We generated a Gaussian random field from a harmonic power spectra that combined both noise and foregrounds. In this work, we separate out the modeling of the CIB and radio fields and explicitly link them to the simulated density fields. We use the \Agora multi-wavelength sky simulation \citepalias{Omori2022Agora} for our CIB modeling approach and the \textsc{WebSky} framework for the radio emission modeling \citepalias{Li:2022:WebskyRadio}. We will detail each component further below.

Once we generate a map of the brightness temperature per frequency, we combine them into a minimum-variance Compton-y map using a standard Linear Combination (LC) approach \citep[\eg][]{Matt:2020:tSZACTDR4, Bleem:2022:tSZ}. See Appendix A2 in \citetalias{Anbajagane:2024:MapBaryonification} for details on the optimal weights used in this LC. While our forward-modeled frequency maps have some changes (e.g., in the CIB and radio modeling as noted below), the weights are still defined as in \citetalias{Anbajagane:2024:MapBaryonification} (see their Figure A1\footnote{While it is true that this work uses slightly different noise properties for the mm-wave data, given our used of the enhanced SO configuration from \citet{Abitbol:2025:SO} rather than the fiducial one of \citet{SO:2019:Forecast} (see Section \ref{sec:mockmaps:surveys}), the impact on the scales we consider in this work is more minimal so Figure A1 in \citetalias{Anbajagane:2024:MapBaryonification} remains a useful reference for the LC weights.}).

We make one note regarding non-Gaussianity from gravitational non-linearities. Our use of the third moments as a summary statistic indicates a clear sensitivity of our results to non-Gaussian information in the mock maps, specifically the three-point correlations in a cosmic field. Similarly, the fourth moments --- of which we use a specific subset (Section \ref{sec:data:stats}) are sensitive to four-point correlations. In the case of WL, the modeling of such correlations is precise as we use the density field from an N-body simulation, which resolves all such non-linearities. For the tSZ, we expect the same to be true as we utilize the simulation's halo catalogs, which contain any N-point correlations in the spatial distribution of halos. Previous analyses of the tSZ bispectrum have found it sufficient to use the halo model approach \citep{Crawford:2014:tSZBispec}, and this is directly analogous to our halo painting method above. Our CIB and radio models are directly tied to the density field and therefore also inherit any N-point correlation function present in it. As discussed in \citet{Crawford:2014:tSZBispec}, the noise-field and the CMB primaries are Gaussian to good approximation.\footnote{It is true that the $\fNL$ signals we focus on here are tied to non-Gaussian features in the CMB primaries. Treating the CMB primary as Gaussian is a simplification in our work, and will require modifications in the future. We note, however, the the CMB is a negligible contribution to the minimum-variance tSZ maps \citep[\eg][]{Raghunathan:2026:SPT}.}

\subsubsection*{Cosmic Infrared Background}

\begin{figure}
    \centering
    \includegraphics[width=\columnwidth]{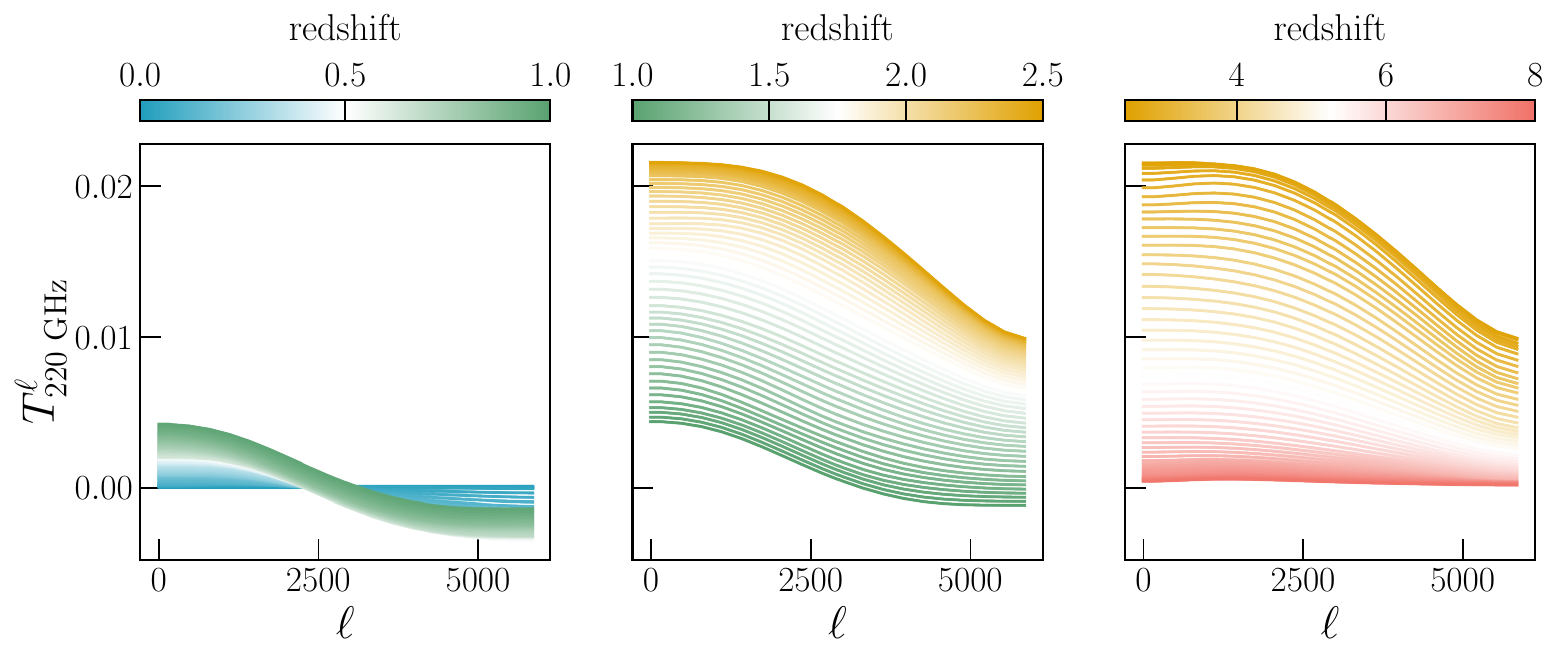}
    \caption{The CIB transfer function (at 220 GHz) for a variety of different redshift slices. The signal is subdominant in the redshift range that overlaps with weak lensing (left panel), and is instead dominant at $z \sim 2.5$. As a result, most of the CIB is uncorrelated with the WL and X-ray probes; see Figure \ref{fig:MapCells} for examples. Note that $T^\ell$ also includes evolution in the galaxy bias and so the peak of this function is shifted to slightly higher redshifts than the $z \approx 2$ peak in the cosmic SFR, as the galaxy bias increases with redshift.}
    \label{fig:CIBTransfer}
\end{figure}

The CIB is sourced by dust in star-forming galaxies that re-radiates stellar light into the infrared wavelengths \citep{Kashlinsky:2005:CIB}. As a result, this emission is highly dependent on the star-formation rates (SFRs) of galaxies over cosmic time. N-body simulations can be used to predict star formation histories (SFHs) for dark-matter halos by utilizing the mass accretion histories (MAHs) of these halos. Such a model is provided in the \textsc{UniverseMachine} framework \citep{Behroozi:2019:UniverseMachine}, which is the method employed in \Agora \citepalias{Omori2022Agora}. The \textsc{Ulagam} simulations are significantly lower resolution than \Agora, and so we cannot follow the same approach in our case. 

\begin{figure}
    \centering
    \includegraphics[width=\columnwidth]{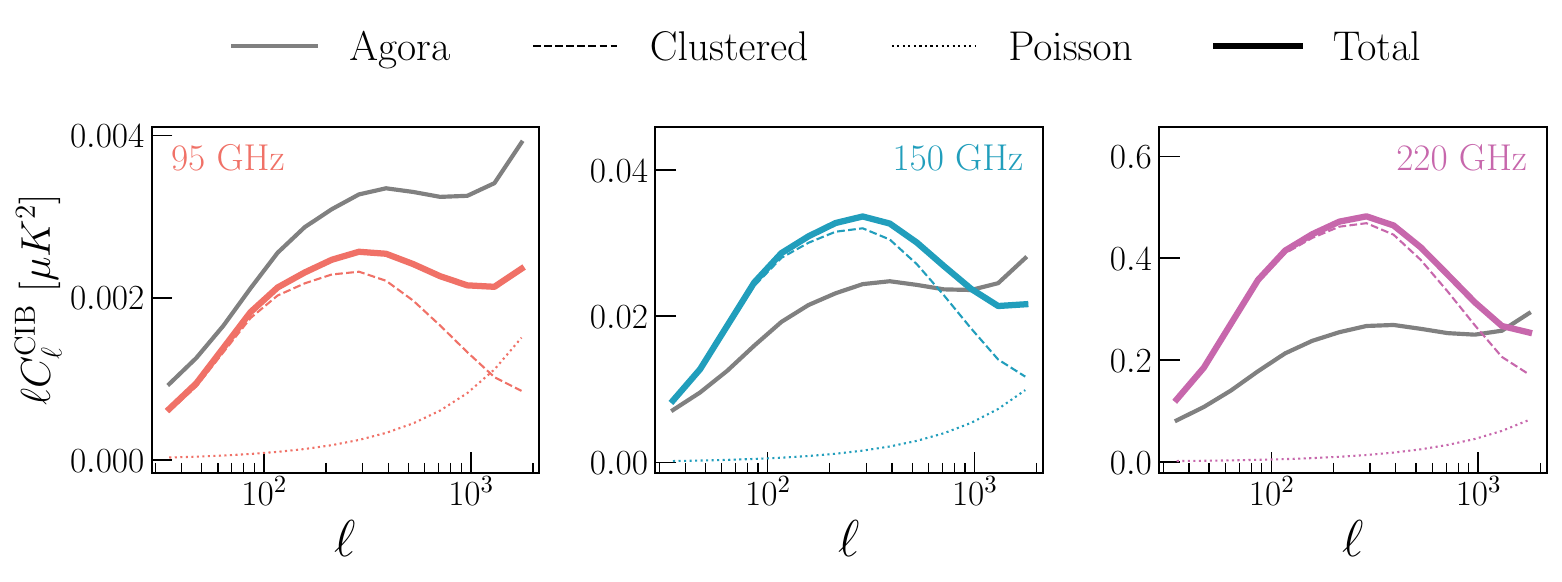}
    \caption{The CIB auto-power spectrum at different frequencies for one full-sky noiseless map from our work (colored lines) and from Agora (gray). We split our signal into the clustered and Poisson components, where the former is modeled as in Equation \eqref{eqn:tSZ:CIB:mainint} while the latter follows the analytic model of \citet{George:2015:foregrounds}. The inclusion of the Poisson term improves accuracy with the Agora prediction. Our model is generally consistent with Agora within 50 to 80\%, which is reasonable given differences in the underlying simulated density fields, the assumed frequency SED, and the approximate nature of our transfer-function method.}
    \label{fig:CIBTest}
\end{figure}

Instead, we use a simpler approach where we learn the relationship between the CIB and density fields from \Agora, and apply it to our simulations. Representing the CIB intensity map in harmonic space as $I_{\rm CIB}^{\ell m}$, we construct its model as
\begin{equation}\label{eqn:tSZ:CIB:mainint}
    I_{\rm CIB}^{\ell m} = \int d\chi T^\ell(\chi) \deltabcm^{\ell m}(\chi)
\end{equation}
where $T^\ell(\chi)$ is a $\ell$-space transfer function estimated from \Agora as,
\begin{equation}\label{eqn:tSZ:CIB:Transfer}
    T^\ell(\chi) = \frac{1}{\Delta \chi} \frac{C^\ell_{I\delta}(\chi)}{C^\ell_{\delta \delta}(\chi)},
\end{equation}
which is the ratio of the cross-correlation between the CIB and density field,\footnote{We use the \Agora density field, which is the DMO density field with no baryonic corrections, i.e. we do not use a $\deltabcm$ from \Agora.} to the auto-correlation of the density. The factor $\frac{1}{\Delta \chi}$ is the width of the shell, and ensures Equation \eqref{eqn:tSZ:CIB:mainint} is a differential quantity. We estimate Equation \eqref{eqn:tSZ:CIB:Transfer} for every redshift shell (or radial distance $\chi$) in \Agora.\footnote{In practice, we construct shells of width $\Delta \chi = 150 \mpc/h$ in \Agora so that adjacent cells have minimal correlations. For highly correlated shells, our simple linearized method in Equation \eqref{eqn:tSZ:CIB:mainint} is not adequate and one would require a matrix of correlations between the different density shells (which is far too expensive, even for maps of $\texttt{NSIDE} = 1024$.). Shells of the required width are obtained by summing every six shells provided in the \Agora public release.} Figure \ref{fig:CIBTransfer} presents this transfer function in differential redshift form, $dT^\ell/dz \propto (dT^\ell/d\chi) / H(z)$, to make it easier to compare to results in the literature. The kernel peaks at $z \approx 2$, consistent with the behavior shown in \citetalias{Omori2022Agora} (see their Figure 19). We can then use Equation \eqref{eqn:tSZ:CIB:mainint} and apply the transfer function on the \textsc{Ulagam} density shells. Figure \ref{fig:CIBTest} shows that our model is within 50\% of the \Agora mocks. Our modeling also includes a free amplitude parameter, $f_{\rm CIB}$, which will capture any uncertainties in the overall normalization of the CIB component.

This method largely recovers the clustering signal present in the CIB. However, there is also a significant Poisson term in the CIB field \citep{George:2015:foregrounds, Reichardt:2021:highell}. This is more dominant at high-$\ell$, and simulation-based models often capture this using halo occupation distribution (HOD) approaches. In this work, we continue modeling the Poisson component of the CIB using the analytic form from \citet{George:2015:foregrounds}.\footnote{Given the Poisson component is dominated by shot noise and not cosmic clustering, it is poorly correlated with the density field, and as a result is not captured well in Equation \eqref{eqn:tSZ:CIB:mainint}. We stress that the \Agora map includes this Poisson component as it explicitly models CIB sources in its modeling.} Figure \ref{fig:CIBTest} shows that our model is similar to the Agora reference maps. We do not expect perfect agreement given differences in simulation resolution, cosmology, and the approximate nature of our Poisson correction.

\subsubsection*{Radio sources}

The auto-correlation of the radio emission is largely dominated by a random Poisson contribution rather than a spatially clustered one. However, the cross-correlation with a density-based tracer can have increased sensitivity to the clustered term. Therefore, we explicitly Poisson sample the radio sources from the density field. For the CIB, this was not necessary as the contribution is already split into clustered and Poisson components, so we explicitly simulated the clustered component and added the Poisson component analytically. Such a split is not done for the radio modeling, so we instead Poisson sample the density field to include the clustering signal, even if it is highly subdominant to the Poisson component. We heavily rely on the approach from \citetalias{Li:2022:WebskyRadio}, used for the \Websky simulations, which itself is based on the model of \citet{Wilman:2008:Radio} and \citet{Sehgal:2010:sims}.

First, the number of radio sources per halo is defined by a HOD relation,
\begin{equation}
    N_{\rm radio}(M) = N_0 (M/M_0)^{0.1}.
\end{equation}
where $N_0$ and $M_0$ are free parameters. For each redshift shell, we determine the number density of radio sources using the HOD scaling (i.e., the number of sources per halo) and the halo mass function. We convert this to a number count per pixel, $N_{\rm 2D}$, by using the width of the shell, and the comoving area of the pixel at the central redshift of the given shell. Second, we sample a set of Poisson sources from the density field through $\text{Poisson}[N_{
\rm 2D}(1 + b_r\deltabcm)]$. The bias $b_r$ is estimated as an average over all sources,
\begin{equation}
    b_r = \frac{\int dM N_{\rm radio}(M) \times \text{HMF}(M) \times b(M)}{\int dM N_{\rm radio}(M) \times \text{HMF}(M)}
\end{equation}
where we use the halo mass function (HMF) and halo bias relation from \citet{Tinker2010Bias}. Having assigned sources to each pixel on the sky, we randomly sample a luminosity for each source using the luminosity function (LF),
\begin{equation}
    p(L) {\rm d\!\ln}L =
    \begin{cases}
        (L/L_b)^m & L < L_b\\
        (L/L_b)^n & L \geq L_b\\
    \end{cases}
\end{equation}
where $L$ is the luminosity at 151 MHz. Finally, we can simply sum the total luminosity per pixel to obtain our final radio frequency map. In practice, we split the radio source sample into two subsets, the Fanaroff-Riley Class I (FRI) and Class II (FRII) radio sources. Each class is found to have significantly different jet/lobe morphology and cosmic evolution \citep{Fanaroff:1974:Radio}. Naturally, the model in \citetalias{Li:2022:WebskyRadio} assigns each class their own HOD parameters, as well as jet/lobe morphology parameters. We use the parameters listed in Table 1 of \citetalias{Li:2022:WebskyRadio}.

\begin{figure}
    \centering
    \includegraphics[width=\columnwidth]{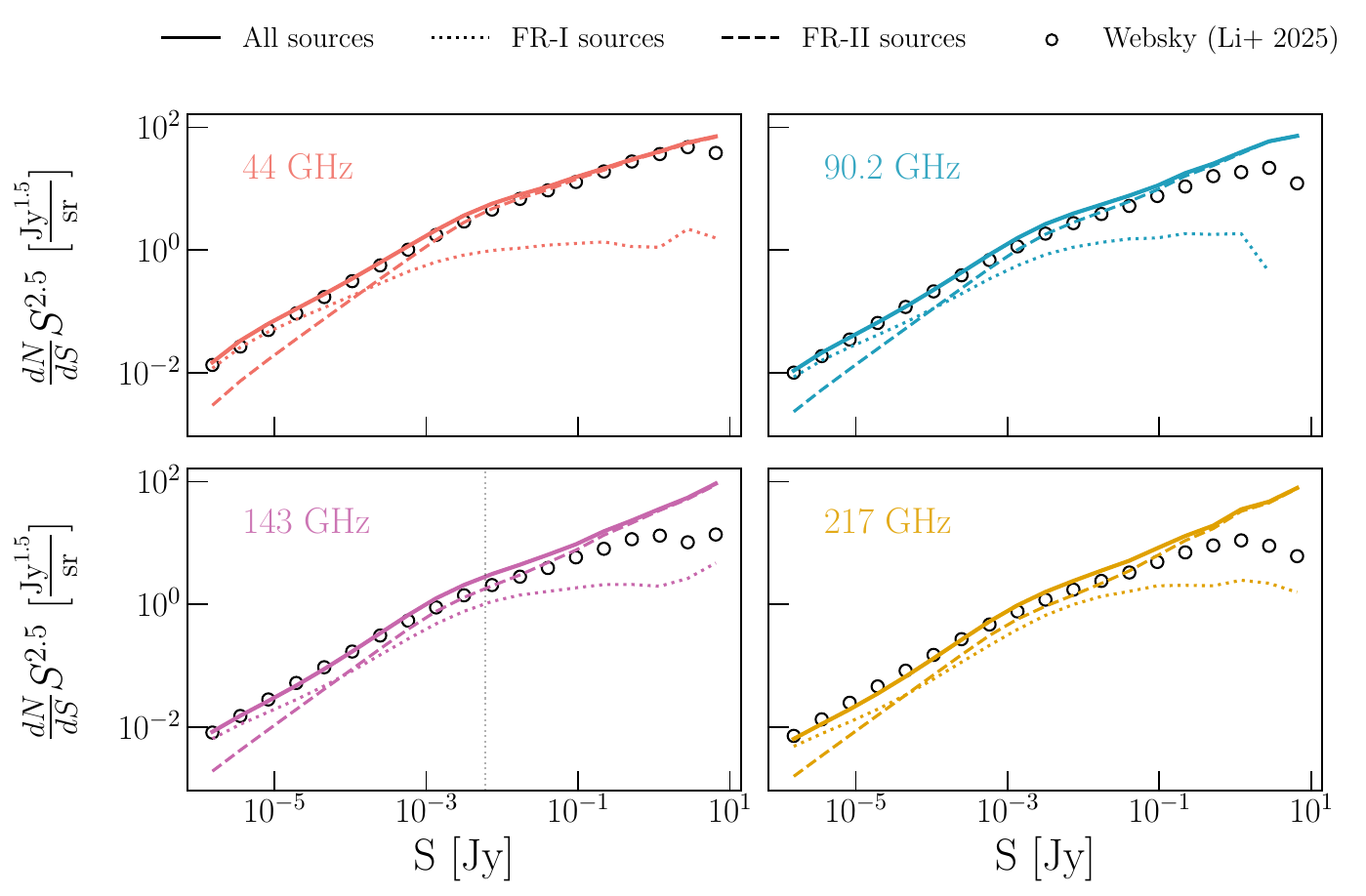}
    \caption{The radio source counts as a function of frequency (panels), split by the two source classes. We show counts from the \Websky reference catalog provided by \citetalias{Li:2022:WebskyRadio}. The luminosity function agrees well, except for some deviations at the high luminosity end, which are expected as we do not use the modified abundance-matching procedure nor the modified SED step from \citetalias{Li:2022:WebskyRadio}. However, such bright sources are masked (the gray line in the 143 GHz panel represents the 6 ${\rm mJy}$ masking threshold above which all sources are masked) and present no concern to our analysis.}
    \label{fig:RadiodNdS}
\end{figure}

The radio source morphology is a key part in the modeling. When drawing the random Poisson sample, $L$ is the total luminosity at 151 MHz. It is then split into jet and lobe components using a constant ratio defined per FR class, and by a relativity beaming factor that is randomly drawn per object; see Equation 8 in \citetalias{Li:2022:WebskyRadio}. The lobe component has a SED of $f^{-0.8}$, where $f$ is frequency. The jet emission is modeled as a cubic power-law as in Equation 12 of \citetalias{Li:2022:WebskyRadio}. The work of \citetalias{Li:2022:WebskyRadio} also presents an abundance-matching and modified SED technique that starts with observed radio point sources. For simplicity, we do not utilize this method here as the correction is predominantly important for bright sources \citepalias[][see their Figure 3]{Li:2022:WebskyRadio} which are anyway masked in the analysis, as we note below. Figure \ref{fig:RadiodNdS} shows the luminosity function of the radio sources we sample, relative to the public \Websky catalogs from \citetalias{Li:2022:WebskyRadio}. There is good agreement, except at the highest luminosities where the abundance matching correction is more important. The gray line in the bottom left panel shows the threshold above which sources are masked, which includes all of the high luminosity objects.

Finally, the above procedure on its own is not adequate for matching the data maps from mm-wave surveys. In particular, the map-making procedure in all CMB surveys explicitly removes point source emission through masking or an equivalent procedure \citep[\eg][]{Naess:2025:ACTDR6, Quan:2026:SPT3G}. The SPT-3G data masks any sources with a flux of $S_{\rm 150 GHz} > 6\,{\rm mJy}$, which are $>16\sigma$ detection in their maps \citep[][see Section III.B.2]{Quan:2026:SPT3G} and the Atacama Cosmology Telescope (ACT) also utilizes similar thresholds \citep[][see their Table 10]{Naess:2025:ACTDR6}. We follow a similar approach here, and simply remove point sources above this threshold from our simulated sample, which effectively masks them from the final maps.\footnote{Such a procedure mimics perfect removal of point-source emission from the data maps. In practice, the point-source masking (plus interpolation) in the data maps is quite clean \citep{Naess:2025:ACTDR6, Quan:2026:SPT3G} and so it is fair to assume point-source residual leakage is minimal for the analysis we discuss here.}

\subsubsection*{Other observational considerations}

In practice, the mm-wave data maps also include a number of additional effects. First, the ground-based maps are filtered in non-trivial ways to avoid contamination from the atmosphere, ground pick-up, etc. \citep[\eg][]{Naess:2025:ACTDR6, Quan:2026:SPT3G}. Second, galactic dust (dominated by neutral hydrogen) both emits and absorbs light in the mm-wave, though the former is the dominant term. The exact relation between the hydrogen density and mm-wave emission is normally fit by regressing a dust-map template against the frequency maps of a survey \citep[\eg][]{Bleem:2022:tSZ}. For simplicity, we ignore both these terms.\footnote{In practice, the latter could be ignored in the forward model if dust cleaning was accurately done on the data maps. The filtering effect must still be accounted for in some fashion as we must model the effect of unrecoverable modes lost during filtering.} Other effects, such as the impact of carbon monoxide (CO) emission could be relevant for the forward models \citep{Kokron:2024:COCMB}, but are currently ignored.

\subsection{X-ray modelling}\label{appx:model:xray}

Thus far, different groups have analyzed the X-ray power spectrum \citep{Lau:2023:XrayPk, Ferreiera:2024:Xray, LaPosta:2025:XraytSZWL}, which is distinct from analyses of X-ray sources like galaxy clusters or AGN. The former analyses have been done primarily through a halo-model approach as detailed in the above works. We follow the modeling guidelines presented in these prior efforts, and translate them into forward models for the map-level effort here.

The key observable in X-ray surveys is the photon count rate per solid angle. X-ray photons are predominantly produced by hot plasma in galaxy clusters, by AGNs, and by X-ray binaries in galaxies. In our work, the latter two are modeled simultaneously under the same prescription. Similar to the case with the thermal SZ, all relevant non-Gaussian three-point correlations are included in our model by basing our modeling prescription on either the simulated halo catalogs or the associated density shells, which exhibit all relevant higher-order correlations. We now discuss our two modeling prescriptions.

\subsubsection*{Galaxy clusters}
The X-ray emission in galaxy clusters is produced by the thermal bremsstrahlung cooling through interactions between electrons and protons \citep{Kravtsov:2012:ClusterReview}. This signal can be represented as,
\begin{equation}\label{eqn:Xray:CRcluster}
    \CR(r) = \int dE \,n_{\rm e}(r) n_{\rm H}(r) \Lambda_{\rm cool}[T(r), Z(r), E(1 + z)] \int dE_{\rm obs} A(E_{\rm obs}) P(E_{\rm obs} | E)
\end{equation}
where $n_{\rm e}$ and $n_{\rm H}$ are the physical number densities of electrons and ionized Hydrogen (or protons), respectively. Then, $T(r)$ and $Z(r)$ are physical temperature and metallicity profiles, and $\Lambda_{\rm cool}$ is the cooling, or photon emissivity, function in units of photons per second per number density squared. We denote the true photon energy as $E$ and the detector-observed energy as $E_{\rm obs}$. Both are observed-frame energies, and we account for photon redshifting by evaluating the cooling function at the rest-frame energy of the source, $E(1 + z)$. The inner integral accounts for the effective area of X-ray detectors as a function of energy, $A(E_{\rm obs})$, and for the energy redistribution between true and measured photon energies, $P(E_{\rm obs} | E)$. Both the energy distribution matrix and the effective area function are publicly available.\footnote{\url{https://erosita.mpe.mpg.de/dr1/eSASS4DR1/eSASS4DR1_arfrmf}} The emissivity function is calculated using the APEC tables provided under the AtomDB package.\footnote{\url{http://www.atomdb.org/download_process?fname=apec_v3_1_3}} The \texttt{Xray.Emissivity} module in \textsc{Vaanam} contains utilities for downloading and constructing these tables, as well as for working with energy response matrices.

The outer integral of Equation \eqref{eqn:Xray:CRcluster} represents the true radial profile of the photon count-rate. This profile is also scaled with redshift as, $\CR(r, z) = \CR(r) \times (1 + z)^{-3}$, to account for cosmic dimming.\footnote{Note that this factor does not exactly match surface brightness dimming, which scales as $(1 + z)^4$, as our observable is a count-rate rather than an energy-rate, and only the latter is impacted by cosmic redshifting.} Equation \eqref{eqn:Xray:CRcluster} also depends on the halo metallicity. Prior works have modeled this as a scale-independent value of $Z = 0.3 \zsun$. In this work, we include the well-established steepening of $Z(r)$ towards the halo core \citep[][see their Figure 3]{Mernier:2018:Metals}, which is also where the X-ray signal peaks given the sharp drop of the term $n_e(r) n_H(r)$ with radius. See Appendix \ref{appx:model:Xray:Z} below for more details. The temperature profile is obtained by computing the thermal gas pressure and dividing by the proton number density,
\begin{align}\label{eqn:temperature}
    T(r) & = P_{\rm th, gas}(r) / n_{\rm p}(r)\\
    n_X(r) & = \frac{\rho_{\rm gas}(r)}{\mu_X m_p}
\end{align}
where $X \in [e, H, \ldots]$ is a given species and $\mu_X$ is the associated molecular weight, $\mu_e = 1.15$ and $\mu_H = 1.32$.

One noteworthy point is the sensitivity of X-ray cluster emission to secondary halo properties. The X-ray cluster population is bimodal in being split between cool-core (CC) and non cool-core (NCC) clusters. These two populations differ in the amplitude of their emission in their cores ($r < 0.05 \Rtwohc$), but their integrated emission is largely regular \citep{Kravtsov:2012:ClusterReview}. We have not included such contributions in our modeling.

\subsubsection*{AGNs and X-ray binaries}

Galaxy clusters are not the sole contributor to the X-ray sky. Two other dominant contributors are AGNs and X-ray binaries. We model both using the data-constrained luminosity functions presented in \citet{Aird:2015:AGNXLF}, in concert with our simulated density shells. There are four separate components --- three AGN models (Unaborbed, Absorbed, and Compton Thick) and the host galaxies of X-ray binaries. We will refer to these as four different source classes.

The procedure here broadly follows that of the radio source modeling used in the mm-wave maps of Appendix \ref{appx:model:tSZ}. For each redshift shell, we estimate the total number density of a given source class and convert it into number counts per pixel, $N_{\rm 2D}$. We then generate a source sample using the sampling function, $\text{Poisson}[N_{\rm 2D}(1 + b_{\rm AGN}\deltabcm)]$. We follow \citetalias{LaPosta:2025:XraytSZWL} in using $b_{\rm AGN} = 1$ for all sources, motivated by \citet[][see their Figure 7]{Comparat:2023:XrayAGN}. We then randomly draw a luminosity per source from the LF, and convert it to a photon count. The LF is given as a double power-law,
\begin{equation}
    \phi(L_X) = K\bigg[\bigg(\frac{L_X}{L_\star}\bigg)^{\gamma_1} + \bigg(\frac{L_X}{L_\star}\bigg)^{\gamma_2}\bigg]^{-1},
\end{equation}
with parameters (including their redshift dependence) taken from Table 9 in \citet{Aird:2015:AGNXLF}. The LF is defined for the [2, 10] keV energy band, and we shift it into the [0.5, 2] keV energy band using an assumed SED for each source class. Following \citetalias{LaPosta:2025:XraytSZWL}, the phenomenological SEDs are informed by \citet[][see their Figure 4]{Aird:2015:AGNXLF}. We also evaluate the LF at the rest-frame energies for the redshift of a given density shell, and then convert them into observed energies at the present time using the relevant SED. For Unabsored and Galaxy classes, the SED is a simple power-law of $dN/dE \propto E^{-1.9}$, while for the absorbed and Comption-thick classes this power-law is multiplied by a sigmoid function that transitions at $E = 4 \,\,{\rm keV}$ and is reduced to 1\% the original value for energies below this scale. Finally, we note that our final estimate of the photon counts per source also includes the effective area and response matrices, as utilized in Equation \eqref{eqn:Xray:CRcluster}.

We do not connect the sampling of X-ray AGN sources here with the radio point source discussed above in the mm-wave modeling. \citet{Pennock:2025:XrayRadio} study radio and X-ray sources in the Bo\"otes and COSMOS field, and do not find a statistically significant correlation between the luminosities in the two regimes. \citet{Georgakkis:2003:Radio} find a detection ($4\sigma$) of X-ray around radio sources, and attribute it to star-forming galaxies rather than AGN. Given we mask all bright radio sources from our mm-wave map models, we expect an even weaker correlation of radio emission and the X-ray AGN. Thus, we ignore such a correlation in our work. Note that this discussion is purely about the sampling of luminosities. The X-ray AGN and the radio populations are still Poisson-sampled from the same underlying density field and thus, are spatially correlated.

Similar to the radio component modeling, we mask all objects above a given brightness. Our threshold for brightness is chosen following \citetalias{LaPosta:2025:XraytSZWL}, who set a threshold of $0.02$ photons per second in the ROSAT data. Converting this to eROSITA DR3 --- since we must account for the different grasps of the surveys \citep{Predehl:2020:eROSITA} --- our threshold is set at $\approx 0.3$ photons per second. All sampled AGN above this threshold are masked.

\subsubsection*{Other observational considerations}
The above discussions present a methodology for generating maps of the photon count rate. Real survey observations include a few other effects. We have already incorporated the primary instrumental effect --- the response of the instrument to an incoming photon of a given energy --- via the inner integral of Equation \eqref{eqn:Xray:CRcluster}. There are two other effects that we also include. The first is the absorption of X-ray photons via the neutral hydrogen in the galaxy. We use neutral hydrogen maps from HI4PI survey \citep{HI4PI2016}, and model the absorption coefficients following \citet{Lau:2023:XrayPk}. Next, we include the exposure time of the eROSITA data,\footnote{Obtained here: \url{https://erosita.mpe.mpg.de/dr1/AllSkySurveyData_dr1/HalfSkyMaps_dr1}} which has large variations with maxima appearing near the poles due to its scanning patterns. The provided maps are for DR1, and we scale these exposure times by a factor of 4.5 to mimic the expected DR3 data. We account for these exposure time variations as,
\begin{equation}
    {\rm CR}_{\rm obs} = {\rm Poisson}({\rm CR}_{\rm true} \times t_{\rm exp}) / t_{\rm exp}.
\end{equation}
So the uncertainty in the count rate varies across the sky according to the exposure time. Point-source masking and background subtraction are two other effects that could be relevant, but these are more relevant for analysis on the real data, compared to Fisher information quantification, so we leave these steps for future studies.

\subsubsection{Metallicity Profiles}\label{appx:model:Xray:Z}

\begin{figure}
    \centering
    \includegraphics[width=\columnwidth]{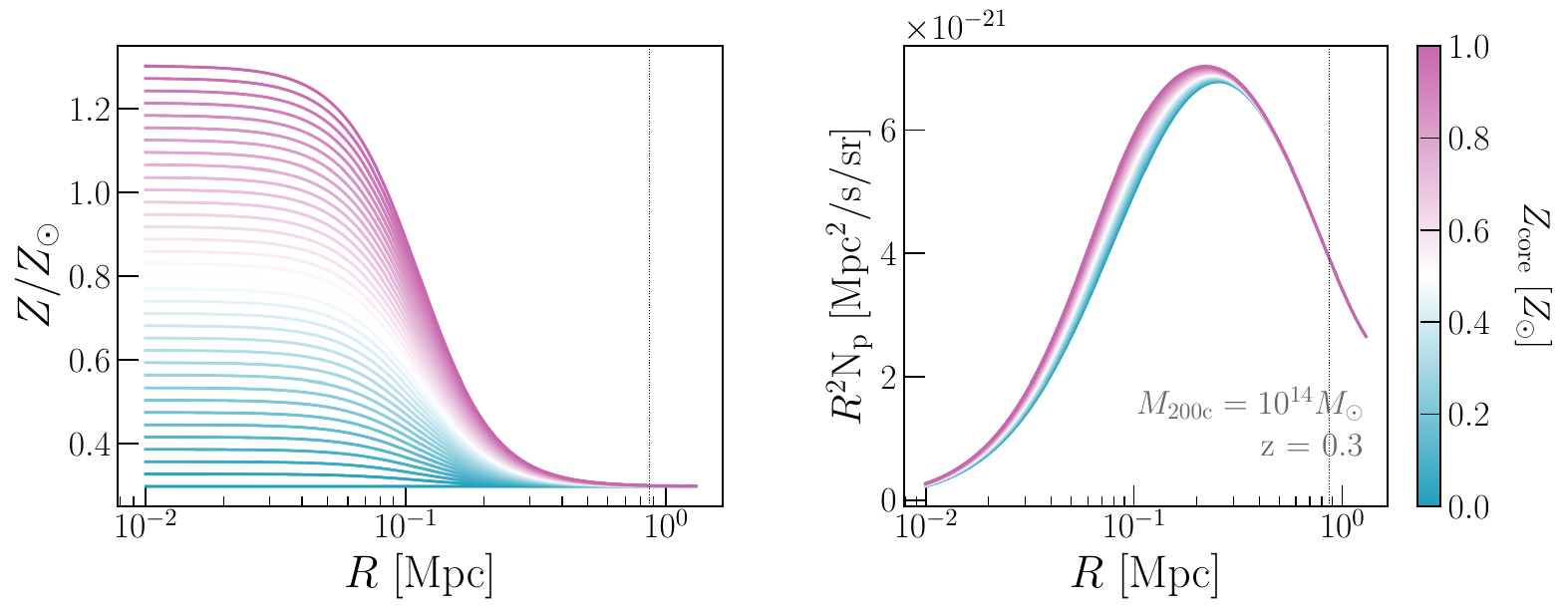}
    \caption{The metallicity profiles (left) and X-ray counts (right) for a single halo of $\Mtwohc = 10^{14} \msun$ at $z = 0.3$, as a function of core metallicity, $Z_{\rm core}$. We scale the counts profile by $R^2$ for visibility but also so that the sum of the profile represents the volume-integrated counts. Current analyses use $Z_{\rm core} = 0$, and the figure shows that varying this parameter has a 10\% impact on the integrated counts on smaller scales, which is of similar magnitude as the AGN contribution to the X-ray sky \citep{McDonald:2026:XrayFlamingo}.}
    \label{fig:Zprofiles}
\end{figure}

In Equation \eqref{eqn:Xray:CRcluster}, we show an explicit dependence on $Z(r)$, the metallicity profile of a galaxy cluster. As mentioned earlier, existing works assume halos have a constant metallicity, $Z = 0.3 \zsun$ \citep{Lau:2023:XrayPk, Ferreiera:2024:Xray, LaPosta:2025:XraytSZWL}. This is broadly motivated by a well-established finding in the X-ray literature, that the gas distributions on larger scales ($R > 0.25 \Rtwohc$) have a constant metal enrichment \citep{Werner:2013:PerseusMetals, Urban:2017:ClusterMetals}. However, it is also established that below this radius, there is a steepening of the metallicity profile \citep[][see their Figure 3]{Mernier:2018:Metals}. We therefore model the metallicity profile as two terms,
\begin{equation}
    Z(r, \Mtwohc, z) = Z_{\rm out}(M, z) + Z_{\rm core}(M, z) \bigg[1 + \bigg(\frac{r}{\theta_{Z_{\rm core}} \Rtwohc}\bigg)^{\gamma_{Z_{\rm core}}}\bigg]^{-1}
\end{equation}
where the normalizations are redshift and mass-dependent as follows, 
\begin{equation}
    Z_{\rm X}(M, z) = Z_{\rm X} (M / M_{\rm Z_{\rm X}})^{\mu_{Z_{\rm X}}} \,\, (1 + z)^{\nu_{Z_{\rm X}}}
\end{equation}
with $X \in \{\text{core}, \text{out}\}$. Using measurements in the literature, we assume fiducial values of $Z_{\rm out} = 0.3 \zsun$,  $\theta_{Z_{\rm core}} = 0.1$, $Z_{\rm core} = 0.8$, $\gamma_{Z_{\rm core}} = 2$, $\mu_{X} = 0$, and $\nu_X = 0$. While we expect there to be some redshift-dependence --- particularly of $Z_{\rm core}$, which could potentially change due to late-time enrichment from stellar evolution in cluster central galaxies \citep{Mernier:2018:Metals} --- we assume this dependence is minor relative to varying the redshift-independent normalization itself. For example, \citet{Mantz:2017:ClusterMetals} find redshift evolution that is consistent with zero within their uncertainties, suggesting at a weaker evolution of the metallicity.

Figure \ref{fig:Zprofiles} shows the impact of assuming a scale-independent metallicity on the X-ray photon counts. The change is of the order 10\% to 20\%, which is at the same level as the AGN contribution to the sky. As a result, we fold $Z_{\rm core}$ into our modeling, and also marginalize over it in our analysis. In practice, we find the constraints on $Z_{\rm core}$ and $Z_{\rm out}$ are broad enough to consider these parameters unconstrained (Figure \ref{fig:results:Contours}). We expect the core term will be more important as we push to smaller scales, by increasing the map resolution to $\texttt{NSIDE} = 2048$ or $\texttt{NSIDE} = 4096$.

\subsection{Summary}\label{appx:model:summary}

\begin{figure}
    \centering
    \includegraphics[width=\columnwidth]{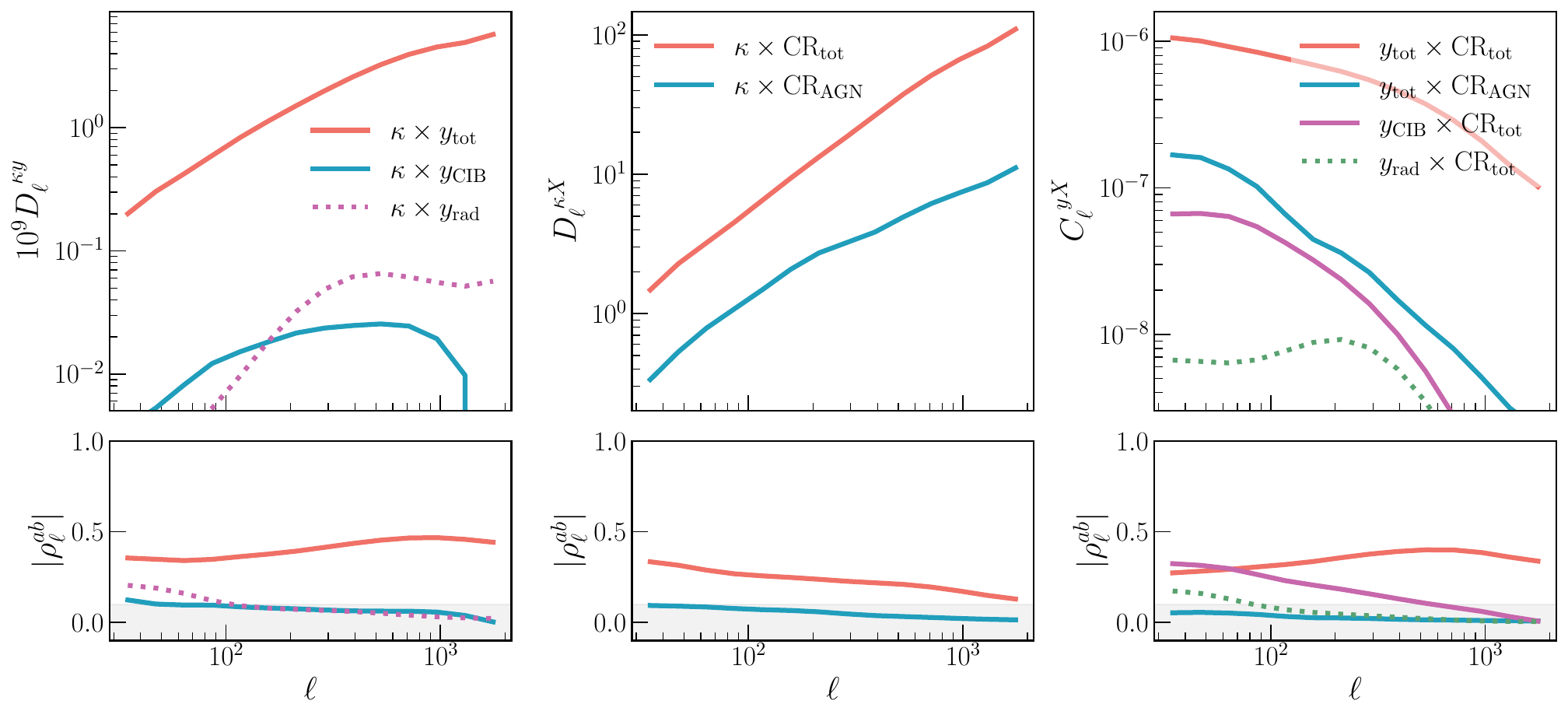}
    \caption{The cross power-spectra (top) and absolute value of the correlation coefficient (bottom) for various maps/components in our forward model, evaluated using one specific simulation. We evaluate these on the noiseless full-sky realizations of the signals. All $y$-map components have been processed via the LC weights, and so can have negative cross-power due to the impact of frequency-dependent weights on components with different SEDs. The radio signal (dotted lines) --- and the CIB on certain scales --- are negative due to this effect. The lensing convergence, $\kappa$, is estimated for the third redshift bin (of five total) defined for the LSST Year 10 analysis, and its lensing signal peaks at $z \sim 0.3$. The light gray band denotes the region $\rho \in [0, 0.1]$.}
    \label{fig:MapCells}
\end{figure}

To summarize, we refer to Figure \ref{fig:MapCells}, which shows an example of the cross-correlations between different components discussed above. We generate a full-sky, noiseless mock dataset from one simulation and compute the harmonic auto/cross-power spectra. The correlations between $\kappa$, $y$, and X-rays are broadly consistent with the best-fit model in \citetalias[][]{LaPosta:2025:XraytSZWL} (see their Figure 2 and Figure 10).\footnote{The X-ray amplitude in our work is roughly $10\times$ higher as the effective detector area of eROSITA is one order of magnitude higher than that of ROSAT, so the photon count-rate increases accordingly. We have also produced these correlations for a ROSAT configuration and find that our predictions match the amplitudes shown in \citetalias{LaPosta:2025:XraytSZWL}.} We also find that the AGN contribution in $\kappa \times \text{X-ray}$ and $y \times \text{X-ray}$ is roughly $\mathcal{O}(10\%)$, once again consistent with the above work. Our X-ray correlations have a slightly larger amplitude, by a factor of 2-3. However, this is expected as the baryon SAM and resulting parameters are different across the two works, and the X-ray observable is highly sensitive to the shape of the inner density profile of the halo gas.

Figure \ref{fig:MapCells} also shows the cross-correlation with various foreground components of the tSZ and X-ray fields. For the case of radio emission, we show the absolute value as the frequency-dependent and $\ell$-dependent weights used tSZ map-making cause the radio emission subcomponent to have an overall negative amplitude due to its SED. The rightmost panel of Figure \ref{fig:MapCells} shows the correlation between the CIB and the X-ray sky is fairly large, contributing at the same level as the AGN correlation. This indicates such terms may also be relevant when analyzing correlations between these two fields. 

Finally, we show the correlation coefficients on the bottom panels, estimated as $\rho^{ab}_\ell = C^{ab}_\ell/\sqrt{C^{aa}_\ell C^{bb}_\ell}$. Most correlations are weak, and at or below $\rho < 0.1$ when correlating with the foreground components. The dominant correlations are between the lensing convergence, the tSZ signal, and the cluster X-ray signal, as expected. The correlation of lensing convergence with CIB is fairly low as the WL signal peaks around $z \sim 0.3$, and only correlates with a small subset of the total CIB emission. The CIB transfer function is quite suppressed in the redshift range best-probed by WL (Figure \ref{fig:CIBTransfer}).

\section{Constraints in $\Lambda$CDM and wCDM}\label{appx:LCDM_wCDM}

\begin{figure}
    \centering
    \includegraphics[width=0.5\columnwidth]{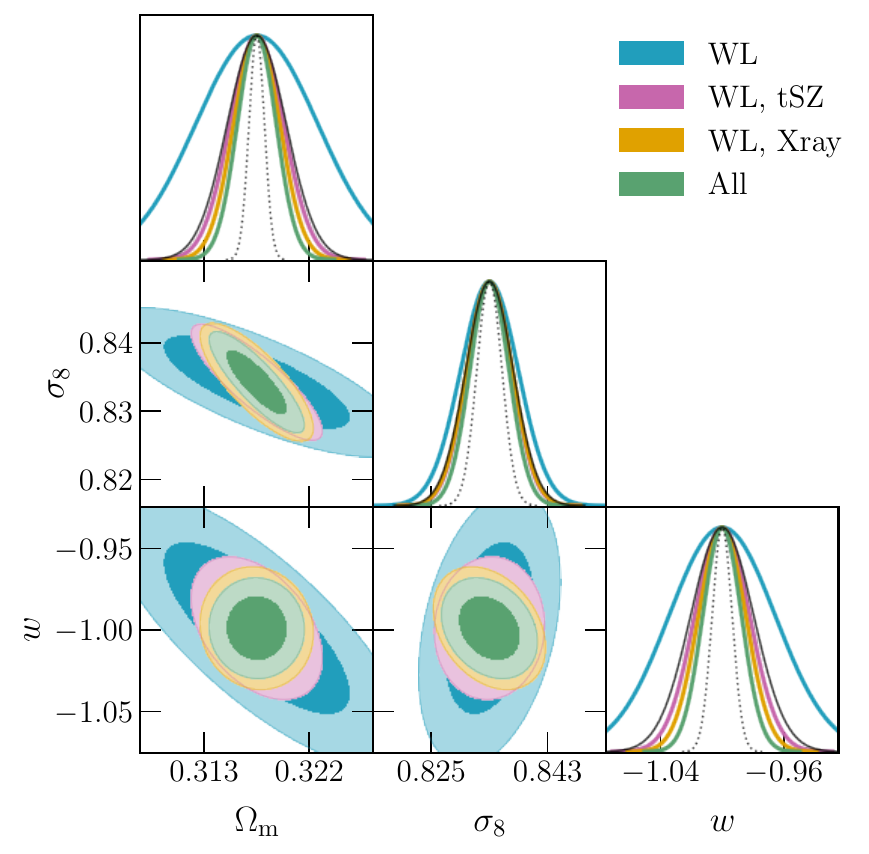}\hspace{20pt}
    \includegraphics[width=0.4\columnwidth]{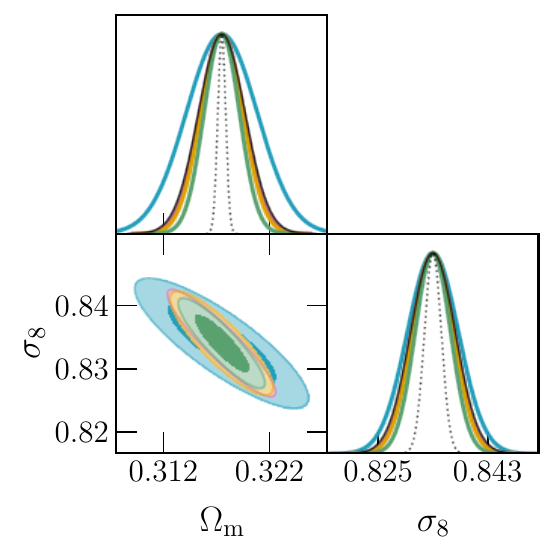}
    \caption{Constraints from a wCDM (left) and LCDM (right) analysis, using the ``All'' combination of probes used in the main analysis. We vary all nuisance parameters, and use the fiducial baryon SAM model. The black solid (dotted) lines show 1D constraints from WL (All data) when fixing all nuisance parameters.}
    \label{fig:wLCDM}
\end{figure}

For completeness, we also explore constraining power from the combination of WL, tSZ, and X-ray on a different extended model; instead of varying $\fNL \neq 0$, we vary the dark energy equation of state, $w \neq -1$. Figure \ref{fig:wLCDM} shows our results, alongside the LCDM results (where we fix $w = -1$) for comparison. In both cases there is a factor of two improvement (relative to WL alone) in constraints, primarily on $\Omega_{\rm m}$. Better constraints on the matter content naturally improve constraints on $w$ given the degeneracy between the two. The combination of probes deliver constraints of $\sigma(w) = 0.015$, while WL-alone provides $\sigma(w) = 0.034$. In LCDM, the improvement in $\Omega_{\rm m}$ continues to be a factor of two.

\section{Numerical convergence}\label{appx:numconvergence}

\begin{figure}
    \centering
    \includegraphics[width=\columnwidth]{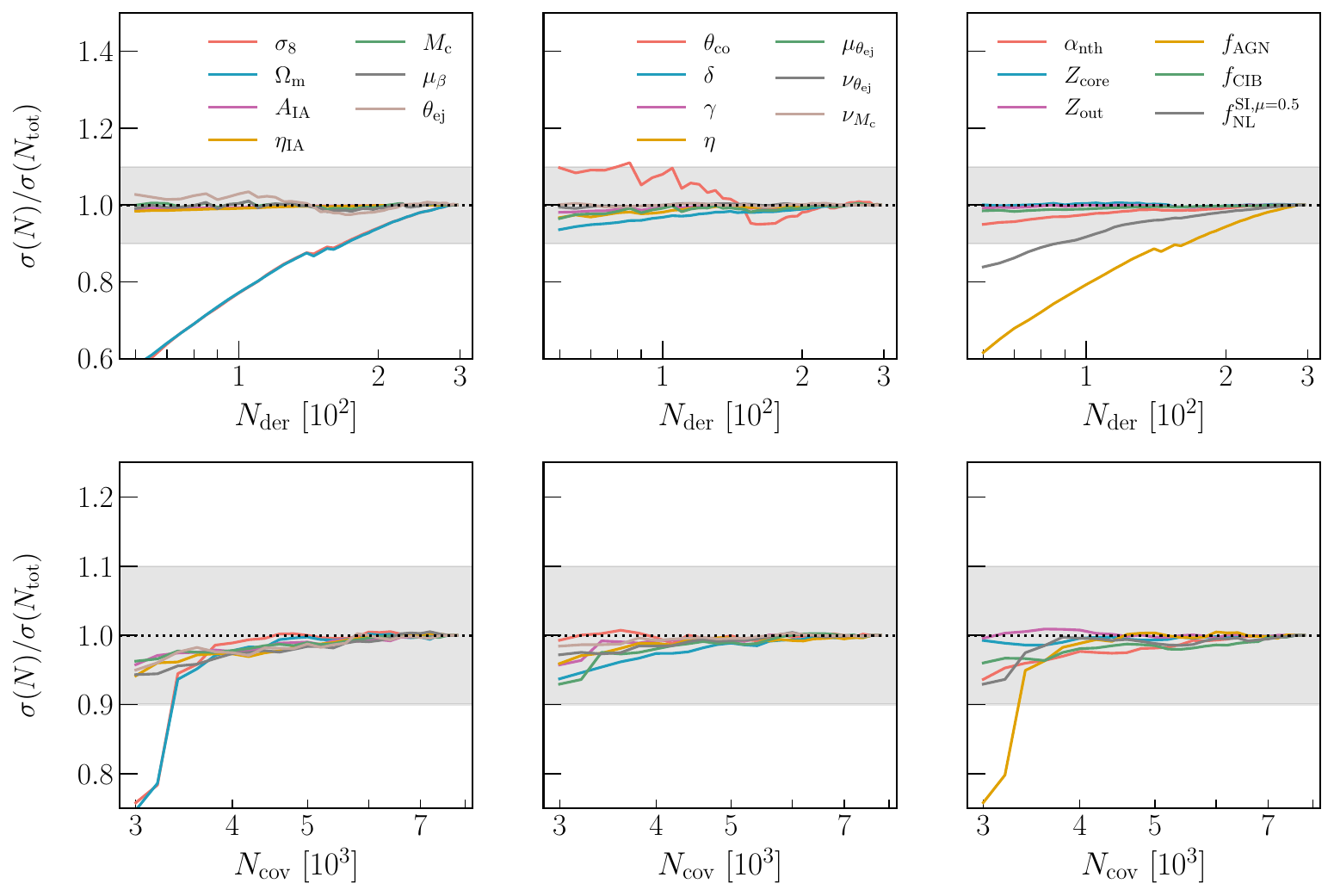}
    \caption{The numerical convergence of our results, estimated by the change in (marginalized) constraining power as we change the number of simulations used to estimate derivatives (top) and covariance (bottom) as defined in Equation \eqref{eqn:Fisher}. All parameters change by less than 10\% if we use half the available simulations.}
    \label{fig:FisherConvergence}
\end{figure}

Numerical convergence is a significant concern in simulation-based Fisher forecasts, as detailed in \citet{Coulton:2023:FisherBias}. We check for its impact by varying the number of simulations used in our estimation of the derivative and covariance. Figure \ref{fig:FisherConvergence} shows the marginalized Fisher constraints on each parameter are converged within 10\% even if we use half the available simulations. We show this for the particular case of the SI model with $\mu = 0.5$, but other model choices provide quantitatively similar results. Most parameters are converged at the 1-2\% level. The current behavior is sufficiently accurate for all qualitative and quantitative discussions in this work.

\section{Fast correlated sampling of Poisson variables}\label{appx:poisson}

When performing Fisher forecasts with simulation-based modeling, one must pay closer attention to the phases of the noise fields used in the forward models. Forecasts require derivatives of the measurement (a set of summary statistics evaluated on a set of maps) with respect to a parameter of interest, $X$. Such derivatives are estimated using simulations produced at parameter values of $X = \mu \pm \delta$, where $\mu$ is the fiducial value and $\delta$ represents the offsets used to compute numerical derivatives. In this case we want the positive ($X^+$) and negative ($X^-$) sims to have highly correlated noise realizations. This way the finite differences of statistics measured in $X^\pm$ are noise suppressed, but the forward models use a statistically accurate representation of the noise. N-body simulation suites are constructed with this choice in mind, as the initial conditions for the $X^\pm$ versions of a given run are generated with the same random seed and therefore are already very correlated \citep[\eg][]{Navarro2020Quijote, Kacprzak2023Cosmogrid, Anbajagane2023Inflation}.

When the noise model is completely independent of the cosmological signal, it is trivial to generated highly correlated noise fields for pairs of simulations, $\text{Sim}^\pm$. We simply decompose the model into $M^\pm = \text{Sim}^\pm + \text{Noise}$. Therefore, we can just generate a Noise term with a single seed and apply it to $\pm$ realizations. This approach is valid for the WL and tSZ models, whose noise components are predominantly Gaussian. In practice, the simulated noise realizations are not perfectly correlated across $M^\pm$ as we still transform the noise in ways that depend on the $\text{Sim}^\pm$ term. For example, the lensing noise model includes source clustering, which depends on the cosmic density field in a given simulation (Appendix \ref{appx:model:WL}). The tSZ model noise is affected by the CIB, which is also sourced by the density field (Appendix \ref{appx:model:tSZ}). However, even under these considerations, the noise field remains highly correlated across the $M^\pm$ realizations.

In the case of X-ray, a similar procedure cannot be performed as the noise is predominantly Poisson in nature. For example, the AGN population is a significant contributor to the X-ray sky noise (even if its spatial correlation signal is smaller than the cluster-dominated term) and our model includes it by Poisson-sampling the density field (Appendix \ref{appx:model:xray}). The observed photon count rate map is also a Poisson-sampled realization of the true count-rate map. In all these cases, we want the Poisson noise in the $M^\pm$ realizations to be highly correlated with each other. This is slightly challenging as the Noise term is fully a function of the cosmic signal (density field, true count-rate field, etc.) as the Poisson rate, $\lambda$, is set by these signals (Appendix \ref{appx:model:xray}). In this Appendix, we detail two specialized methods that enable correlated sampling of Poisson variables in a computationally efficient manner.

\subsection*{D.1\hspace{10pt}    Fast inverse sampling}\label{appx:poisson:invcdf}

Pseudo-random number generators (PRNGs) start from a random state (a seed) and deterministically produce a stream of numbers that approximate random sequences.\footnote{Although not truly random, their long periods and statistical properties make them effectively indistinguishable from random sequences for almost all practical purposes \citep{Matsumoto:1998:MersenneTwister}.} The \texttt{numpy.random.poisson} method uses rejection sampling, to draw Poisson samples \citep{Hormann1993, Knuth1997}. As a result, when sampling $N$ random Poisson variables, we cannot reliably determine which number(s) from the random stream are used to generate the $i^{\rm th}$ random variate.

\begin{figure}
    \centering
    \includegraphics[width=\columnwidth]{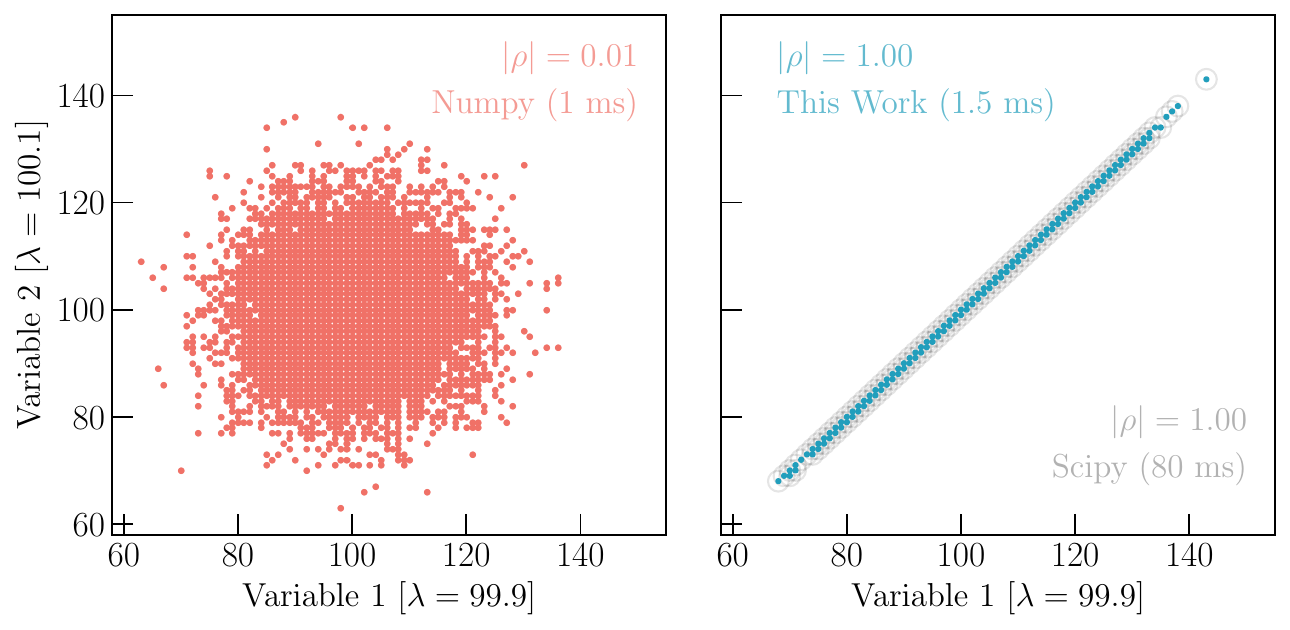}
    \caption{A demonstration of the limitations of the \texttt{Numpy} Poisson-sampling routines for generating correlated samples. The two sets of $10^4$ numbers, ``Variable 1'' and ``Variable 2'', are generated with the same random seeds but are drawn with Poisson rates, $\lambda$, that differ by 0.2\%. This shift is enough to completely decorrelate the samples as \texttt{Numpy} uses rejection sampling and is therefore highly stochastic across different input rates, $\lambda$. Using an inverse sampling technique, which is a deterministic sampling, generates correlated variables as expected. Our method (Appendix \ref{appx:poisson:invcdf}) is vastly faster than the \texttt{Scipy} implementation while still enabling correlated sampling. Compute times are listed in parentheses, in units of milliseconds.}
    \label{fig:Poisson}
\end{figure}

To illustrate this more concretely, we describe the following example using \texttt{Numpy}. Let us initialize a random state with a given seed, $s$, and generate $N$ numbers with a Poisson rate, $\lambda_A$. We then initialize a new random state with the same seed, $s$, and generate another set of $N$ numbers but with a different rate, $\lambda_B$. The two sets of produced numbers, $A$ and $B$, are essentially uncorrelated ($\rho \lesssim 0.01$, where $\rho$ is the Pearson correlation coefficient) even for cases with $\lambda_A / \lambda_B = 0.99$, i.e., the two rates are consistent at 1\%. This is precisely because the $i^{th}$ number in set A and the same in set B use different sections of the random number stream and therefore quickly become uncorrelated. While the random stream is the same for both sets (since we set the same seed in the random state), the section of the random stream used to draw the $i^{\rm th}$ number varies dramatically due to the choice of rejection sampling in \texttt{Numpy}. Thus, it is impossible to generate highly correlated Poisson samples when relying on the computationally efficient techniques used in \texttt{numpy.random.poisson}. Figure \ref{fig:Poisson} shows a demonstration of this issue, alongside the resolution that we now detail below.

This issue can be trivially resolved by using inverse sampling --- we generate a set of random uniform numbers, $u \in U[0, 1]$, and then transform them into a Poisson variable using the inverse cumulative probability, $\text{CDF}^{-1}(\lambda, u)$ --- instead of rejection sampling. The same set of uniform random variates, $u$, can be generated in both simulations $\text{M}^\pm$, and can then be transformed into Poisson variable using the appropriate $\lambda$ rates of each simulation. We find this approach results in the Poisson random variates of the $X^\pm$ simulated realizations being correlated at $\rho > 0.95$.\footnote{We will never get a perfect correlation as the noise field depends on the density/count-rate field and the latter are still different (even if mostly similar) across the $X^\pm$ simulations.} Such an implementation is available in \texttt{Scipy}, but is computationally expensive (the runtime for a $\text{NSIDE} = 1024$ map is 90 times slower than \texttt{Numpy}) as the inverse CDF method evaluates the inverse of a Gamma function via a root-finding proecedure for each random variate it samples.

In this work, we use a simple technique starting from the fact that the Poisson CDF can be written as,
\begin{align}\label{eqn:PoissonCDF}
    \text{CDF}(k)  & = \exp[-\lambda] \times \sum_{j = 0}^{\floor{k}}\frac{\lambda ^ j}{j!}\\[7pt] 
    & = \exp[-\lambda] \times \Upsilon(k)\nonumber
\end{align}
where $\floor{k}$ is the floor of $k$. We abbreviate the summation as the function $\Upsilon(k)$ to compress notation for the discussion to follow. Our notation also makes implicit the dependence of this function on $\lambda$. For a given random number, $u \in [0, 1]$, and a fixed rate $\lambda$, we evaluate the CDF at increasing integer values of $k$ until $\text{CDF}(k) \geq u$. The first value of $k$ that satisfies this condition is returned as the random variate.\footnote{The use of $\text{CDF}(k) \geq u$ instead of $\text{CDF}(k) = u$ is because the Poisson CDF is still discrete and therefore cannot satisfy exact equalities.} For $k < 100$, the CDF can be evaluated rapidly as it requires only simple summations. Notice the recursion relation,
\begin{equation}
    \Upsilon(k) = \Upsilon(k - 1) +  \frac{\lambda^{\floor{k}}}{\floor{k}!}.
\end{equation}
So we can start from $k = 0$ and increment the value until we satisfy the condition, $\text{CDF}(k) \geq u$. As a result, this method is just as fast as the \texttt{Numpy} approach noted above, but with the attractive ability to produce correlated samples via inverse sampling.

In the limit of large $\lambda \gg 1$, the above sum-based approach will suffer from numerical precision limitations. A large $\lambda$ requires that the sum extend to large $k$. In these cases, we will find instances of $\Upsilon(k) \approx \Upsilon(k + 1) \Rightarrow \Upsilon(k) = \Upsilon(k + 1)$ because the difference in the two numbers is below the numerical precision. As a result, we set two conditions in our approach: (i) the sum can be evaluated to a maximum of $k = 10^4$, and if we still cannot satisfy $\text{CDF}(k) \geq u$ then we terminate the sampling and report a failure; and (ii) if $\Upsilon(k)$ and $\Upsilon(k+1)$ are the same number under \texttt{float64} precision, then the sampling is terminated and reported as a failure. For all such failed samples (which are indicated by a sentinel value), we perform inverse-sampling using the \texttt{Scipy} root-based method instead. For our use-case, the failure rate is at the percent level, so the overall speedup (relative to the \texttt{Scipy}-only case) is still a factor of 50. For a fixed set of uniform variates $u$, we have confirmed our approach produces numerically identical samples as the \texttt{Scipy} method. Our implementation is provided as the \texttt{utils.poisson_sample_fast} method in the \textsc{Vaanam} package.

\subsection*{D.2\hspace{10pt} Key-based uniform sampling}

The method of the above section allows us to assign random variates such that the Poisson random variate in Pixel $i$ of simulation A is highly correlated to that in Pixel $i$ of simulation B. In the case of our AGN and Radio models, we have two stages of Poisson sampling --- first we Poisson sample a number of objects per pixel and then sample a luminosity per object in that pixel. The number of objects per pixel will not be identical between simulations A and B. As a result, if we draw a stream of random numbers and partition them for the objects in both simulations, the $N^{\rm th}$ object in pixel $i$ will have different random numbers in simulation A compared to simulation B, and therefore different luminosities. As a result, even though the number of objects per pixel is highly similar, the Poisson noise in the sampled luminosities (and therefore the final emission maps) will be essentially uncorrelated.

We resolve this by making the random number a function of a few input properties, rather than generating a random stream of numbers and partitioning it appropriately between the objects in different pixels. Our random variate is now obtained as,
\begin{equation}
    u = f({\tt seed, pixind, objind}).
\end{equation}
where the tuple $({\tt seed, pixind, objind})$ is constructed from a random seed, the pixel index, and the object index in that pixel. In this setup, once we fix the seed, the $N^{\rm th}$ object in the $i^{\rm th}$ pixel always draws the same random variate from the function, $f$. For the choice of this function, $f$, we use a python-based implementation of the \texttt{SplitMix64} algorithm, which takes a 64-bit number and generates a random number from it. While its statistical properties are less robust compared to the Mersenne-Twister algorithm \citep{Matsumoto:1998:MersenneTwister}, it is still perfectly adequate for our work. Our implementation is provided as the \texttt{utils.keyed\_uniform} method in \textsc{Vaanam}.
%%%%%%%%%%%%%%%%%%%%%%%%%%%%%%%%%%%%%%%%%%%%%%%%%%

% Don't change these lines
\label{lastpage}
\end{document}